\documentclass[apj]{emulateapj}

\pdfoutput=1

\shorttitle{Slicing the Monoceros Overdensity}
\shortauthors{CONN ET AL.}

\begin{document}


\title{Slicing The Monoceros Overdensity With SUPRIME-CAM}

%

\author{Blair C. Conn\altaffilmark{1}, Noelia E. D. No\"{e}l\altaffilmark{1,2}, Hans-Walter Rix\altaffilmark{1}, Richard R. Lane\altaffilmark{3},  Geraint F. Lewis\altaffilmark{4}, Mike J. Irwin\altaffilmark{5}, Nicolas F. Martin\altaffilmark{6}, Rodrigo A. Ibata\altaffilmark{6}, Andrew Dolphin\altaffilmark{7} and Scott Chapman\altaffilmark{5}}
\altaffiltext{1}{Max Planck Institut f\"{u}r Astronomie, K\"{o}nigstuhl 17, 69117, Heidelberg, Germany}
\altaffiltext{2}{ETH Z\"{u}rich, Institute for Astronomy, Wolfgang-Pauli-Strasse 27, Building HIT, Floor J, CH-8093 Zurich, Switzerland}
\altaffiltext{3}{Departamento de Astronom\'{i}a, Universidad de Concepci\'on, Casilla 160-C, Concepci\'on, Chile}
\altaffiltext{4}{Sydney Institute for Astronomy, School of Physics, The University of Sydney, A28, Sydney, 2006, Australia}
\altaffiltext{5}{Institute of Astronomy, Madingley Road, Cambridge, CB3 0HA, U.K.}
\altaffiltext{6}{Observatoire de Strasbourg, 11 Rue de l'Universit\'e, F-67000, Strasbourg, France}
\altaffiltext{7}{Raytheon Company, PO Box 11337, Tucson, AZ, 85734-1337, United States of America}




\begin{abstract}
We derive distance, density and metallicity distribution of the stellar Monoceros Overdensity (MO) in the outer Milky Way, based on deep imaging with the Subaru Telescope. We applied CMD fitting techniques in three stripes at galactic longitudes: {\it l} $\sim$ 130$\arcdeg$, 150$\arcdeg$, 170$\arcdeg$; and galactic latitudes: +15$^{\circ}\leq${\it b}$\leq$+25$^\circ$. 

The MO appears as a wall of stars at a heliocentric distance of $\sim$10.1$\pm$0.5 kpc across the observed longitude range with no distance change. The MO stars are more metal rich ([Fe/H]$\sim -$1.0) than the nearby stars at the same latitude. These data are used to test three different models for the origin of the MO: a perturbed disc model, which predicts a significant drop in density adjacent to the MO that is not seen; a basic flared disc model, which can give a good match to the density profile but the MO metallicity implies the disc is too metal rich to source the MO stars; and a tidal stream model, which, from the literature, bracket the distances and densities we derive for the MO, suggesting that a model can be found that would fully fit the MO data. Further data and modeling will be required to confirm or rule out the MO feature as a stream or as a flaring of the disc.
\end{abstract}





\section{Introduction}

The Monoceros Overdensity (MO) is an extensive stellar structure found in the outer regions of the Milky Way at Galactocentric distances of $\sim$15 - 18 kpc. It was originally discovered in the Sloan Digital Sky Survey (SDSS) by \citet{2002ApJ...569..245N} and subsequent observations reveal a similar structure in many directions around the Galaxy \citep{2003ApJ...588..824Y,2003MNRAS.340L..21I,2003ApJ...594L.119C,2005MNRAS.362..475C,2005MNRAS.364L..13C,2006MNRAS.367L..69M,2007MNRAS.376..939C,2008MNRAS.390.1388C,2008AJ....135.2013C,2011ApJ...730L...6S}. From this, it has been concluded that it forms a coherent structure from Galactic longitudes of {\it l} = $60^\circ$ - $300^\circ$ and straddles both sides of the Galactic plane. While the approximate extent of the MO is tentatively mapped, its origins are somewhat obscure.

The MO formation scenarios fall into three broad categories: tidal tails from an accreting dwarf galaxy \citep{2004MNRAS.348...12M, 2005ApJ...626..128P}; misidentification of normal Galactic warp/flare profiles \citep{2004A&A...421L..29M,2006MNRAS.368L..77M,2006A&A...451..515M,2007A&A...472L..47L, 2011A&A...527A...6H}; and perturbed disc scenarios whereby a close encounter with a massive satellite induces rings in the outer disc from local material  \citep{2008ApJ...688..254K,2011MNRAS.414L...1M,2011MNRAS.tmp.1860G,2011Natur.477..301P}.  

If we consider each scenario briefly then the first scenario is a Galactic accretion event, where the MO is envisioned to be the tidal tails of a dwarf galaxy merging in-Plane with the disc. Such a scenario is attractive since it links with the $\Lambda$-Cold-Dark-Matter cosmology where galaxies form via successive accretion events  \citep{1978MNRAS.183..341W} and the discovery of many stellar streams in the halo of the Milky Way \citep{2002ApJ...569..245N,
2006ApJ...642L.137B,2006ApJ...651L..29G,2009ApJ...693.1118G}. The proposed progenitor for the MO is the putative Canis Major dwarf galaxy, first discussed in \citet{2004MNRAS.348...12M}. If true, the MO could be a relic of formation as discussed in \citet{2002ARA&A..40..487F}.

The second scenario is used to explain both the stellar overdensity in Canis Major and the MO in terms of standard properties of a galactic disc, that is, the warp and the flare. The Milky Way disc is observed, with various tracers, to warp up in the first two quadrants and warp down in the second two quadrants. For instance, \citet{2002A&A...394..883L} follows this feature with red clump giant stars while \citet{2004mim..proc..165Y} uses pulsars as tracers. As the density of the disc drops with increasing galactic radii, it thickens and flares. The Canis Major dwarf galaxy is therefore the disc of the Galaxy dipping below the plane and the MO is the intersection of the flaring disc at high latitudes above the plane.

Finally, the third scenario invokes the interaction between a dark matter satellite and disc to induce the formation of rings at large galactic radii. The repeated passage of these satellites through the disc drives the formation of spiral arms and rings. This has been tested for low inclination satellite encounters \citep{2008ApJ...688..254K} and more recently for the Sagittarius dwarf galaxy ~\citep{2011MNRAS.414L...1M,2011MNRAS.tmp.1860G,2011Natur.477..301P}, which is on a polar orbit. 

All of these scenarios are able, with varying levels of success, to fit the general spatial and kinematic profiles of the MO and previous attempts to find some decisive evidence or prediction has not ruled out any of these possibilities. More and more observations of the MO, both photometric and spectroscopic are becoming available (SEGUE\footnote{Sloan Extension for Galactic Understanding and Exploration}, PanSTARRS\footnote{Panoramic Survey Telescope And Rapid Response System}, SkyMapper, etc) and so these degeneracies may be broken soon. 

In order to shed light into the above dilemma, we secured observations using the Subaru telescope. Our goal is to investigate the spatial density profile of the MO and also to compare our observational results with the different theoretical scenarios presented in the literature. In Section~\ref{analysis} we discuss the data preparation in terms of the observations, reduction and calibration of the dataset. Section~\ref{secDATA} outlines the analysis of the color magnitude diagrams using CMD fitting techniques. In Section~\ref{discus} we compare our findings with the current scenarios of formation for the MO and in Section~\ref{conclusion} we present our conclusions.

\section{Data Preparation}\label{analysis}

\subsection{Observations and Reduction}\label{obsaat}

\begin{figure*}
\includegraphics[width=120mm, angle=270]{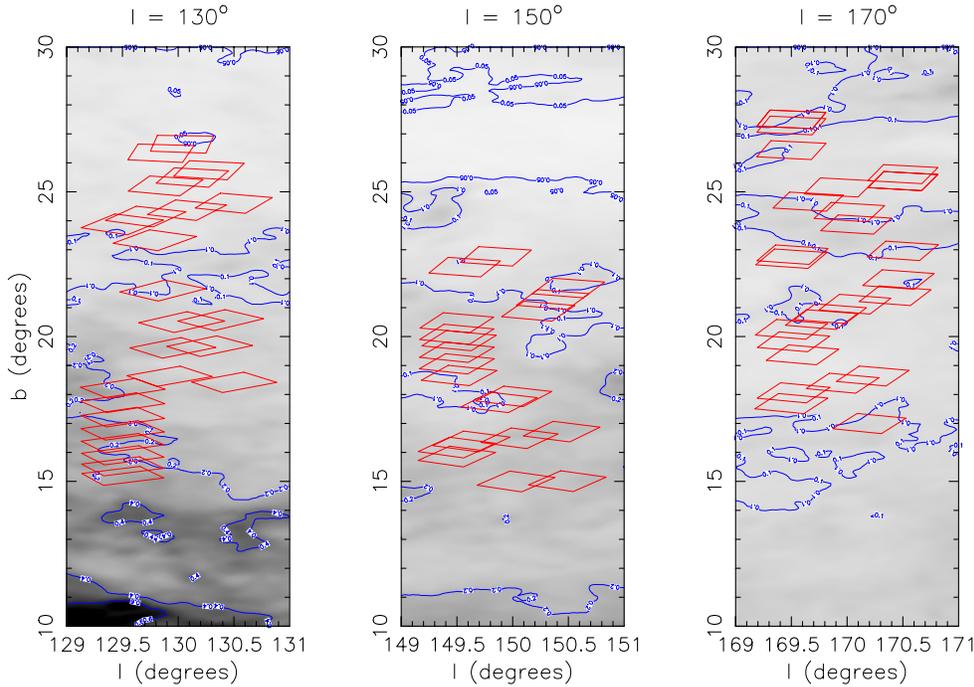}
\caption{Survey layout against Galactic Extinction from the dust
  maps from ~\citet{1998ApJ...500..525S}. The red polygons represent the position of the fields surveyed here. The final survey locations are the result of observational constraints and data quality issues. The grey scale shows the underlying dust extinction with the blue contours outlining the $E(B-V)$ values of 0.05, 0.1, 0.2, 0.4 and 0.6.}\label{figsurvey}
\end{figure*}

The data were collected using the SUPRIME-CAM Wide Field Imager on the Subaru Telescope in Hawaii. SUPRIME-CAM is a 10 chip camera with a field of view of 34\arcmin $\times$ 27\arcmin and a pixel scale of 0$\arcsec$.20. These observations took place in Service mode and were carried out on the 9$^{\rm th}$ of November 2007 and the 1$^{\rm st}$ of January 2008. These observations are summarized in Table~\ref{ObsTable}. Roughly, 180 frames were taken in two filters, Sloan $g$ and $r$, and arranged into the 3 stripes across the thick disc of the Galaxy. Each $g$ frame was 124 seconds and each $r$ frame was 76 seconds. Figure~\ref{figsurvey} shows the survey layout for each stripe with the location of the fields depicted as red polygons overplotted on the local dust extinction contours. Implementing the program in this manner with SUPRIME-CAM, allows for deep observations to be obtained across large areas in a short period of time as required by this study. The final survey locations are the result of observational constraints and data quality issues.

The data were reduced using the Cambridge Astronomical Survey Unit Wide Field Camera Pipeline \citep{2001NewAR..45..105I}. This pipeline was originally developed for the Isaac Newton Telescope Wide Field Camera and has since been modified to work on most of the Wide Field Imagers available today. The pipeline reduced data has been bias-subtracted, flat fielded using twilight sky flats and then flat fielded again using a dark sky super-flat. Following this, the photometry and astrometry have been determined using the same pipeline. The accuracy and completeness of the photometry will be discussed in Section~\ref{completeness}. The astrometry, based on the 2MASS point source catalogue, is typically accurate to between 0.2 and 0.3 arc seconds.

\begin{deluxetable}{lcccc}
\tabletypesize{\scriptsize}
\tablecaption{Summary of Observations.\label{ObsTable}}
\tablewidth{0pt}
\tablehead{
\colhead{Region}& \colhead{Fields}&\colhead{Date Observed}&\colhead{Filter} &\colhead{Median Seeing}
}
\startdata
130 stripe &  1 - 16 & 2007-11-09 & {\it g,r} &  0.71\arcsec,0.62\arcsec\\
  \nodata & 20 - 32 & 2008-01-08 & {\it g}  & 0.78\arcsec\\
 \nodata & 17 - 29 & 2008-01-08 & {\it r}  & 0.69\arcsec\\
150 stripe &  1 - 32 & 2007-11-09 & {\it g} & 0.59\arcsec\\
 \nodata&  1 - 32 & 2008-01-08 & {\it r} & 0.74\arcsec\\
170 stripe &  1 - 32 & 2007-11-09 &  {\it g,r}   &  0.47\arcsec,0.5\arcsec\\
\enddata	
\end{deluxetable}

\subsection{Correcting the Photometry using SDSS\label{sdss}}

Having generated the catalogue of sources and classifying them, the photometric calibration was performed by cross-matching sources with the SDSS Data Release 6 ~\citep{2008ApJS..175..297A}. At the time of the survey only a few of the fields overlapped with the SDSS and so the calibration was performed on those fields and then applied to the others according to their date of observation. Two offsets were needed as it was noticed that Chip 10 has a much lower efficiency than the rest of the array. Since the 150 stripe also had SDSS corrections available, when correcting the other fields the choice between these two was based on which night those observations were taken on. The final photometric solution has a typical scatter of 3.5\% in the $g$-band and 2.4\% in the $r$-band around the SDSS values. 

\subsection{Correcting the Photometry using dust extinction maps}

\begin{figure*}
\includegraphics[width=110mm,angle=270]{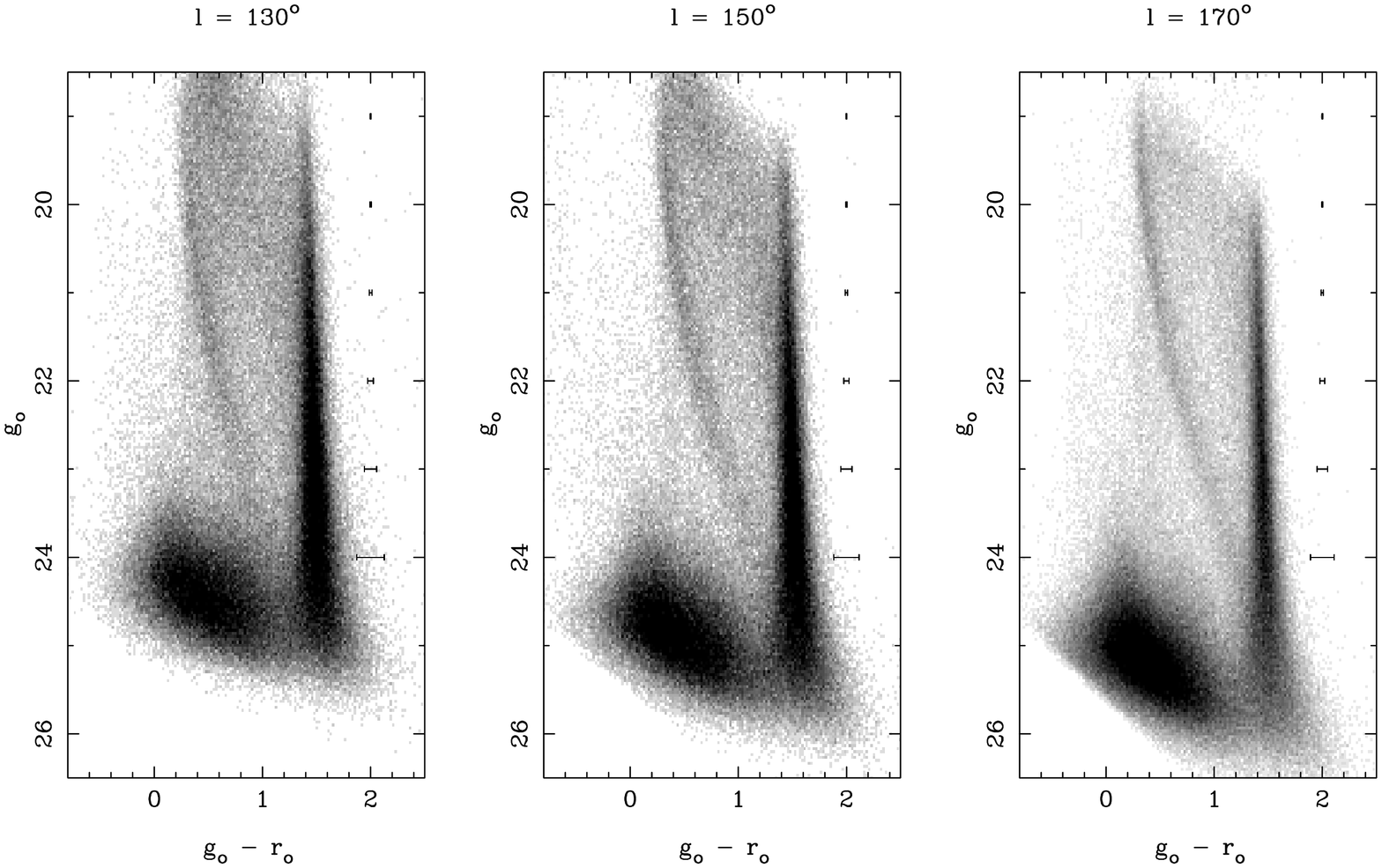}
\caption{Deep Hess CMDs of point sources obtained for each of the three stripes. The data has been extinction corrected according to \citet{1998ApJ...500..525S} and ~\citet{2000AJ....120.2065B}. The cloud of sources in the bottom-left of each panel stems from unresolved galaxies. A difference in the image quality between the observations of {\it l} = 130$^\circ$ and the {\it l} = 170$^\circ$ stripe is responsible for the increase in depth of the resulting CMDs (see Table~\ref{ObsTable}). Main sequence turn-off (MSTO) stars, which we associated with the MO, are found at $g_{o} - r_{o} \sim 0.4, g_{o}\sim 19.5$. The strong overdensity of stars at $g_{o} - r_{o} \sim 1.5$ are the local low mass dwarf stars.}\label{figfinalphot}
\end{figure*}

After the initial correction using SDSS, a correction based on the
dust maps of ~\citet{1998ApJ...500..525S} combined with the adjustment of
~\citet{2000AJ....120.2065B} was implemented.  The final photometry is presented as Hess CMDs in Figure~\ref{figfinalphot}. In general, the dust contamination is less than an $E(B-V)$ of 0.2, for the majority of the survey. The data here has been extinction corrected and represents all the fields for each stripe combined into one figure. 

\subsection{Magnitude Completeness}\label{completeness}
\begin{figure}
\epsscale{.45}
\includegraphics[width=60mm,angle=270]{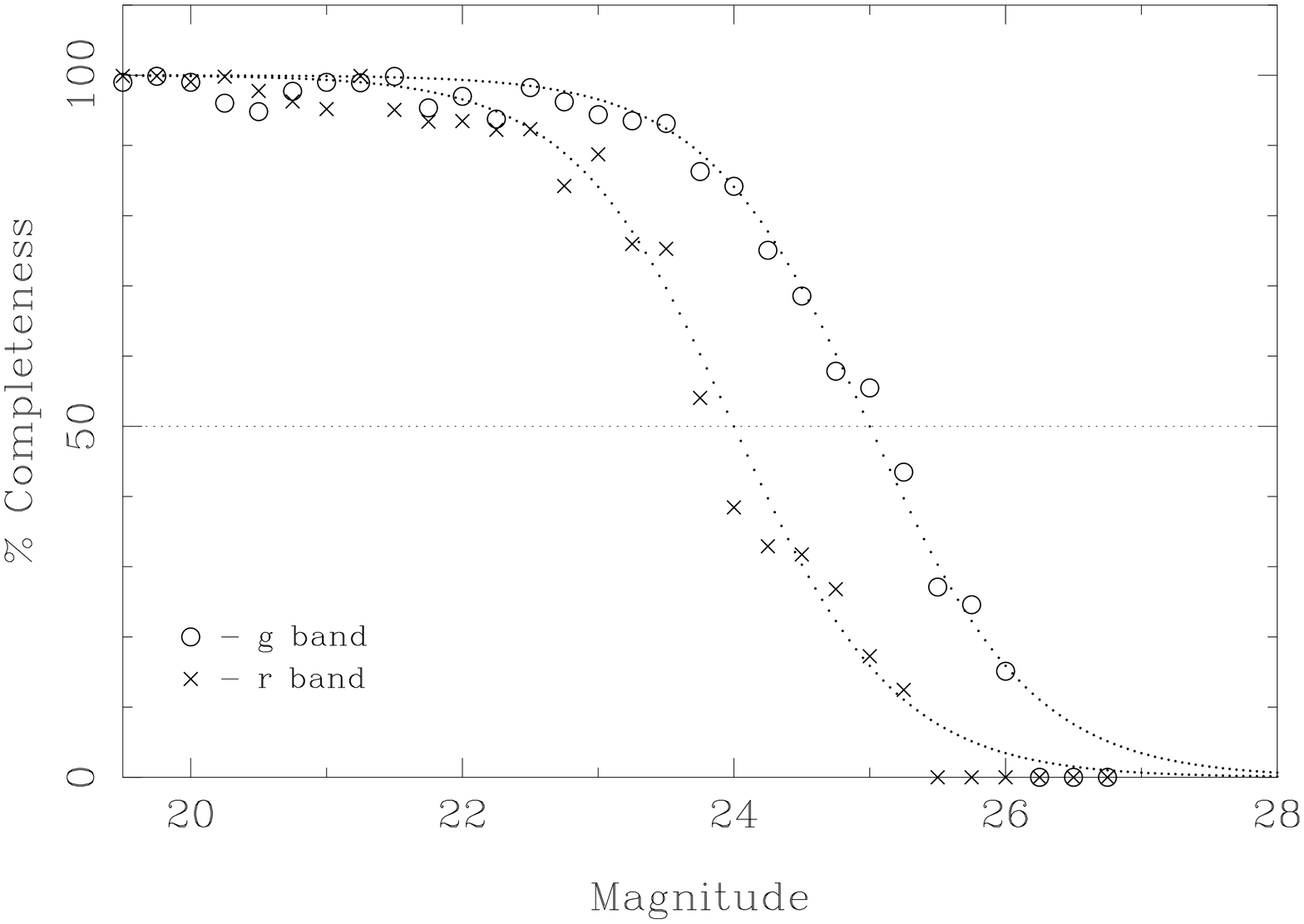}
\caption{Example completeness data and fit for a field in the {\it l} = 150$^\circ$ stripe using the stars in the pointings which overlap. The completeness is determined by whether a star detected in either frame is  recovered in the other frame. This is then binned by magnitude and presented as the percentage completeness of recovered divided by detected. Circles for the $g$-band objects and crosses for the $r$-band objects.}\label{figcomp}
\end{figure}

For those fields which overlap with each other, the completeness of
this sample has been determined in the same manner as used
in the 2MASS All-Sky Point Source Catalogue
\citep{2006AJ....131.1163S}.  By determining the fraction of stars
that are detected in both overlapping images as a function of
magnitude with respect to the total number of stars observed, an
estimate of the completeness can be made. This photometric
completeness curve is then fit by the Logistic function:
\begin{equation}
  CF = \frac{1}{(1 + e^{(m - m_c)/ \lambda})}
  \label{eqncomplete}
\end{equation}
where m is the magnitude of the star, m$_c$ is the magnitude at 50\%
completeness and $\lambda$ characterizes the width of the rollover from 100\% to
0\% completeness. The average values found in each field are presented
in Table~\ref{CompTable} and an example completeness profile is shown in Figure~\ref{figcomp}. 

\begin{deluxetable}{lccc}
\tabletypesize{\scriptsize}
\tablecaption{Mean Magnitude at the 50\% completeness level ($m_c$) and typical width of the rollover function ($\lambda$) for each filter. See Equation~\ref{eqncomplete}.\label{CompTable}}
\tablewidth{0pt}
\tablehead{
\colhead{Region} & \colhead{$m_c$ ($g_\circ$)} & \colhead{$m_c$ ($r_\circ$)} & \colhead{$\lambda$}
}
\startdata
130 Stripe & 24.2 & 23.2 & 0.73\\
150 Stripe & 24.8 & 23.8 & 0.55\\
170 Stripe & 25.0 & 24.2 & 0.70\\
\enddata
\end{deluxetable}

\subsection{Survey Field CMDs}\label{Survey Fields}
The pipeline provides a set of classifications for different objects. We are interested in those classified as stellar and possibly stellar objects.  
The detailed information on the categories is given in \citet{2001NewAR..45..105I}. In short, each processed frame in the pipeline is analyzed using an object detection algorithm based on \citet{1985MNRAS.214..575I,1997ilt..book...35I}. Generated parameters include information on position, intensity and shape. 
To discriminate between background galaxies and real objects three flux estimates are made: integration of the flux above the specific age; the detection isophote for each image is expanded using an elliptical aperture to perform a curve-of-growth analysis; and a 'poor man's' PSF fit using a radius equal to the FWHM. The stellar objects and possibly stellar objects are then selected from all the fields within a given stripe and plotted in a single CMD. Figure~\ref{figfinalphot} shows the three ($g_o$-$r_o$,$g_o$) deep Hess CMDs with each containing approximately 200,000 stars which correspond to the three longitudinal stripes of the survey. To allow the low density MO feature to be clearly visible, the CMD greyscale is scaled using the square root of the counts shown in each pixel. These exceptional CMDs, reaching more than three magnitudes below the oldest MS turnoffs ($g_\circ \sim 19.5$), are the deepest observations of the MO to date. The CMDs show the old MS population along its complete extent from the blue turn-off region to faint red MS stars. The high quality of the photometry and the small errors on the main sequence allow us to secure the distance-metallicity degeneracy in the CMDs. This ensures we can quantitatively measure the stellar content of the MO.

\section{Data Analysis}\label{secDATA}
\subsection{CMD Fitting Technique}\label{sectMATCH}
In order to obtain the stellar populations' structure at the location of the Monoceros Overdensity we used the MATCH software package ~\citep{2002MNRAS.332...91D} in its distance-fitting mode. MATCH was originally developed to obtain quantitative star formation histories (SFHs) and age-metallicity relations for systems in which all the stars are assumed to be at the same distance. For this purpose, the distance is fixed and the age and metallicity are independent variables. In this paper, we apply the CMD-fitting techniques to span the local stellar populations within the Milky Way. For this goal, we can no longer consider that all stars are equidistant and, thus, the distance is a free parameter. In order to limit the number of free parameters, we define a set of template stellar populations for comparison with the data. 

In the same manner as explained in \citet{2010ApJ...714..663D} we used the SDSS $g$ and $r$ isochrones provided by \citet{2004A&A...422..205G}. Given that both the thick disc and stellar halo are known to have old stellar populations, we considered a fixed age range at $\sim$13 Gyrs (10.1$<$ log[t/yrs]$<$ 10.2), 30\% binary fraction, a Salpeter Initial Mass Function ~\citep{1955ApJ...121..161S}, and three metallicity bins, sufficient to describe the halo and thick disc: [Fe/H]= $-$0.7, [Fe/H]= $-$1.3,  [Fe/H]= $-$2.2 [see ~\citet{2010ApJ...714..663D} for more details]. Stars with intermediate metallicities are inferred from the relative weight of these three template populations. The thin disc stars are avoided as they have a broader range of ages and metallicities which would make it difficult to disentangle from a combination of only three different metallicity templates. To avoid edge effects, the model templates were created for distance moduli between 7 ($\sim$250 pc) and 22 ($\sim$250 kpc) in steps of 0.2.

The basis of MATCH is the Hess diagram, a binned CMD in which the value of each bin is the square root of the number of stars.  Synthetic Hess diagrams are then created for a range of ages and metallicities initially assuming 1 M$_{\odot}$/yr star formation rate (SFR) which is then scaled and combined to best match the observed diagram. The synthetic diagrams are convolved with the photometric errors and completeness profile of the data to provide a realistic comparison with the data. When comparing the observed CMD with the synthetic CMD, MATCH uses a Poisson Maximum Likelihood statistic to determine the best-fitting single model or linear combination of models.

\begin{figure*}
\centering
\includegraphics[width=150mm, angle=0]{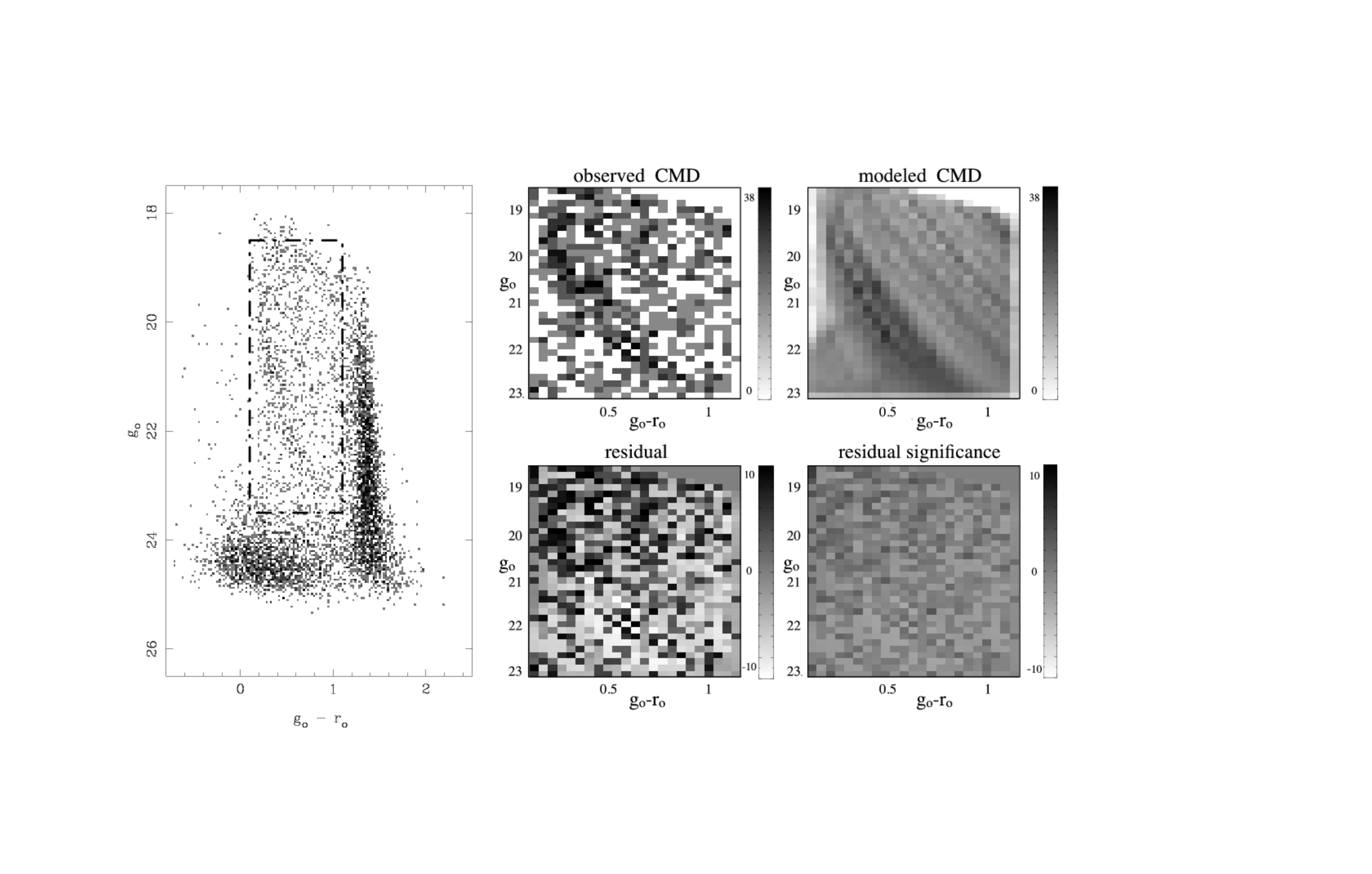}
\caption{The left hand figure shows an observed CMD for a single frame of the stripe {\it l} = 130$^\circ$. The long dashed rectangle represents the region used in our analysis. The four right hand figures show
 the output of the CMD fitting routine MATCH (see Section~\ref{sectMATCH}). The upper panels show the observed CMD (left) and best-fit model to the observed CMD (right). 
 The lower panels are residuals after subtracting the model from the observed CMD (left) and residual significance (right) based on a fit parameter including the number of stars in the observed and model CMD. The grey scale bar represents the number of stars in the Hess diagram bins.
  The analysis was made in the regions $18.5<g_o<23.0$, $18.0<r_o<23.0$ and $0.1<g_o-r_o<1.1$ of the CMDs shown in Figure~\ref{figfinalphot}. In this way, we
avoid the clump at $g_o>23.0$ and $g_o-r_o<1.0$ mainly due to unresolved background
galaxies and local dwarf stars located at $(g_o-r_o)>1.5$. As seen in Figure~\ref{figcomp} we have 95\% completeness at $g_o=23.0$ and 90\% at $r_o=23.0$.}\label{figmodelres}
\end{figure*}

We used stars in the magnitude range 18.5 $< g_{o} <$ 23.0  and 18.0 $< r_{o} <$ 23.0 and in the color range 0.1 $< g_{o} - r_{o} <$ 1.1 (Figure~\ref{figmodelres}). These color cuts ensure that we do not include faint, red stars belonging to the thin disc, while the magnitude cuts secure that we do not include spurious objects such as misclassified galaxies. For each population template, MATCH provides the star formation rate in M$_{\odot}$/yr for each distance modulus bin. The star formation rates are then converted into stellar mass density. 

Figure~\ref{figmodelres} shows an example of an observed CMD and its best-fit model CMD for a single frame in stripe ${\it l} =130^\circ$. For each single frame one Hess diagram is created. The observed CMD is shown on the left hand side; the region
 used in our analysis is depicted with the dot-dashed rectangle. The four panels on the right hand side are: The observed Hess CMD (top left), the model Hess CMD (top right), the residual Hess CMD after subtracting the model from the data (bottom left) and the residual significance, based on the number of stars expected in the model Hess diagram (bottom right). The model CMD reproduces well the main features of the observed CMD, such as the main plume of old MSTO stars, especially since we assumed only a simple model population.

\begin{figure}
\includegraphics[width=80mm, angle=0]{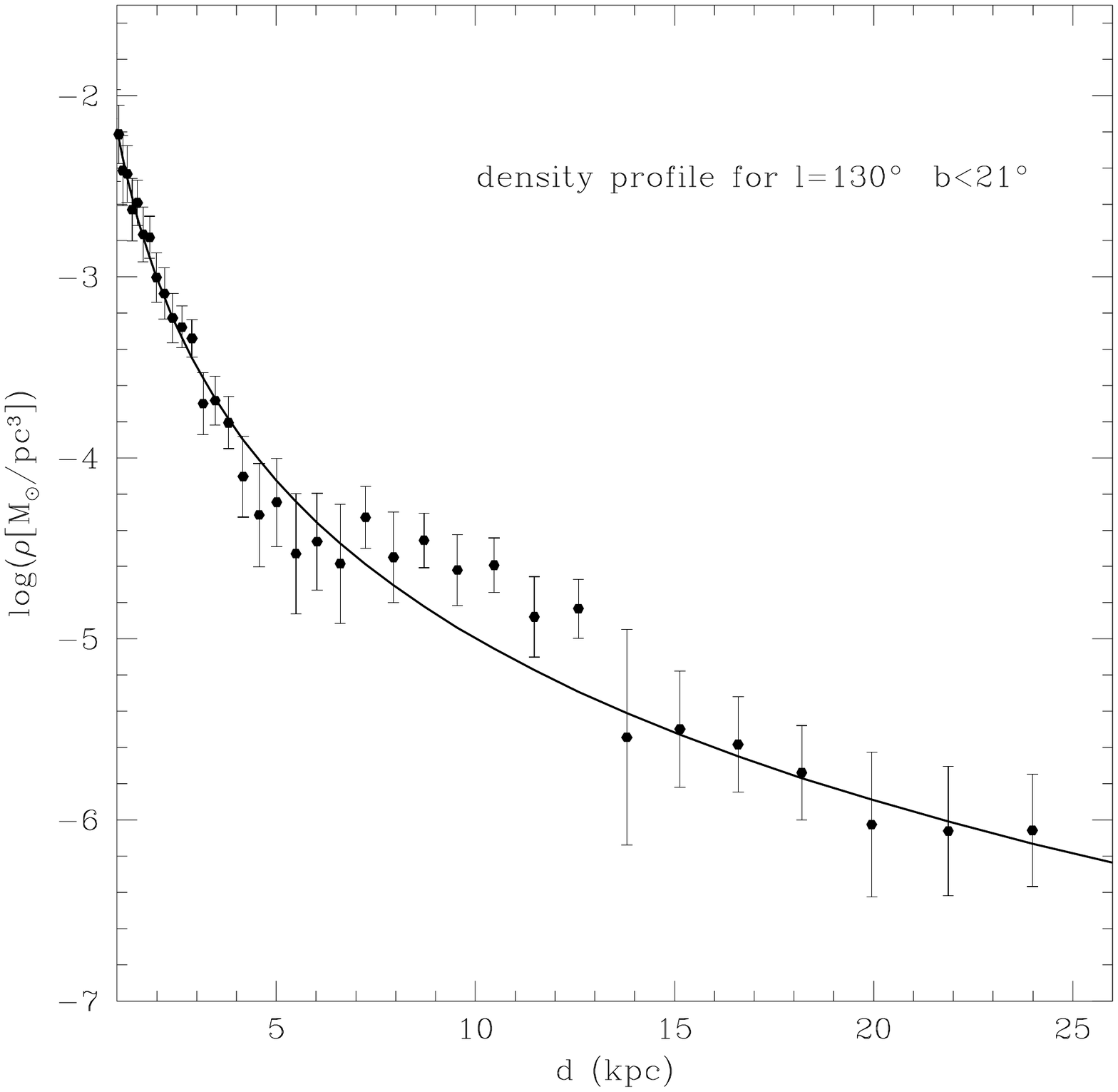}
\caption{Mean stellar mass density distribution for the stripe with {\it l} = 130$^\circ$, {\it b} $<$ 21$^\circ$. The stellar density decreases with increasing distance with a conspicuous deviation at around 10 kpc from the Sun, corresponding to the location of the MO. A S\'{e}rsic profile is fitted to the density distribution to remove the Milky Way background and allow for a cleaner detection of the MO. Error bars were calculated from the uncertainty associated to the density of each bin from the Monte Carlo tests. This is achieved by multiply resampling the stars within the CMD space and re-running MATCH on each resampled population.
\label{figsersic}}
\end{figure}

\subsection{Density and Metallicity gradients}\label{densitygrads}

\begin{deluxetable*}{lccccccc}
\tabletypesize{\scriptsize}
\tablecaption{Results from the MATCH analysis, Distance, depth and density. \label{ResultsTable}}
\tablewidth{170mm}
\tablehead{
\colhead{Region} & \colhead{Mean latitude} & \colhead{Distance Heliocentric} & \colhead{Depth of MO} &\colhead{Stellar Number Density}&\colhead{S\'{e}rsic Index / }&\colhead{S\'{e}rsic Scale}\\	
\colhead{} & \colhead{($b^\circ$)}&\colhead{(kpc)}&\colhead{(kpc)}&\colhead{(counts)}&\colhead{reduced $\chi^{2}$}&\colhead{ Length (kpc)}
}
\startdata
130 Stripe (upper) & 24 & 10.4$\pm$0.5 & 3.8$\pm$0.4& 463&7.0/1.4&0.2\\
130 Stripe (lower) & 18 & 10.7$\pm$0.4 & 3.7$\pm$0.4&671&8.0/1.2&0.1\\
150 Stripe (upper)& 21 & 9.3$\pm$0.5 & 1.8$\pm$0.2&553&13.0/1.2&0.11\\
150 Stripe (lower)& 17 & 9.6$\pm$0.2 & 3.1$\pm$0.4 & 609&6.0/1.5&0.54\\
170 Stripe (upper)& 25 & 10.2$\pm$0.3 & 2.5$\pm$0.3&427&5.15/1.5&2.43\\
170 Stripe (lower)& 20 & 10.5$\pm$0.2 & 1.8$\pm$0.2&588&3.55/1.6&0.47\\
\enddata
\end{deluxetable*}

MATCH provides the star formation rate (SFR, in M$_\odot$/yr) corresponding to each population template for each distance modulus bin and this is transformed into a stellar mass density. Figure~\ref{figsersic} shows the density profiles for the lower half of the {\it l} = 130$^\circ$ stripe plotted against heliocentric distance. A S\'{e}rsic profile is then fit to the underlying stellar population and is shown overplotted on the data. The stellar density shows a clear decrease with distance and includes a conspicuous deviation in the distance range $7 <$ d (kpc)$ < 13$. This ``excess'' in the density distribution, present in all our fields, is due to the MO. The steep exponential profile in the inner $\sim$5 kpc is due to the contribution of the thick disc population. From $\sim$5 kpc onwards the inner halo component dominates up to $\sim$20 kpc when the outer halo begins to rule, producing a flattening in the density profile. This corresponds well to the density profiles reported in ~\citet{2010ApJ...714..663D}. 

\begin{figure}
\begin{center}
\includegraphics[width=80mm, angle=0]{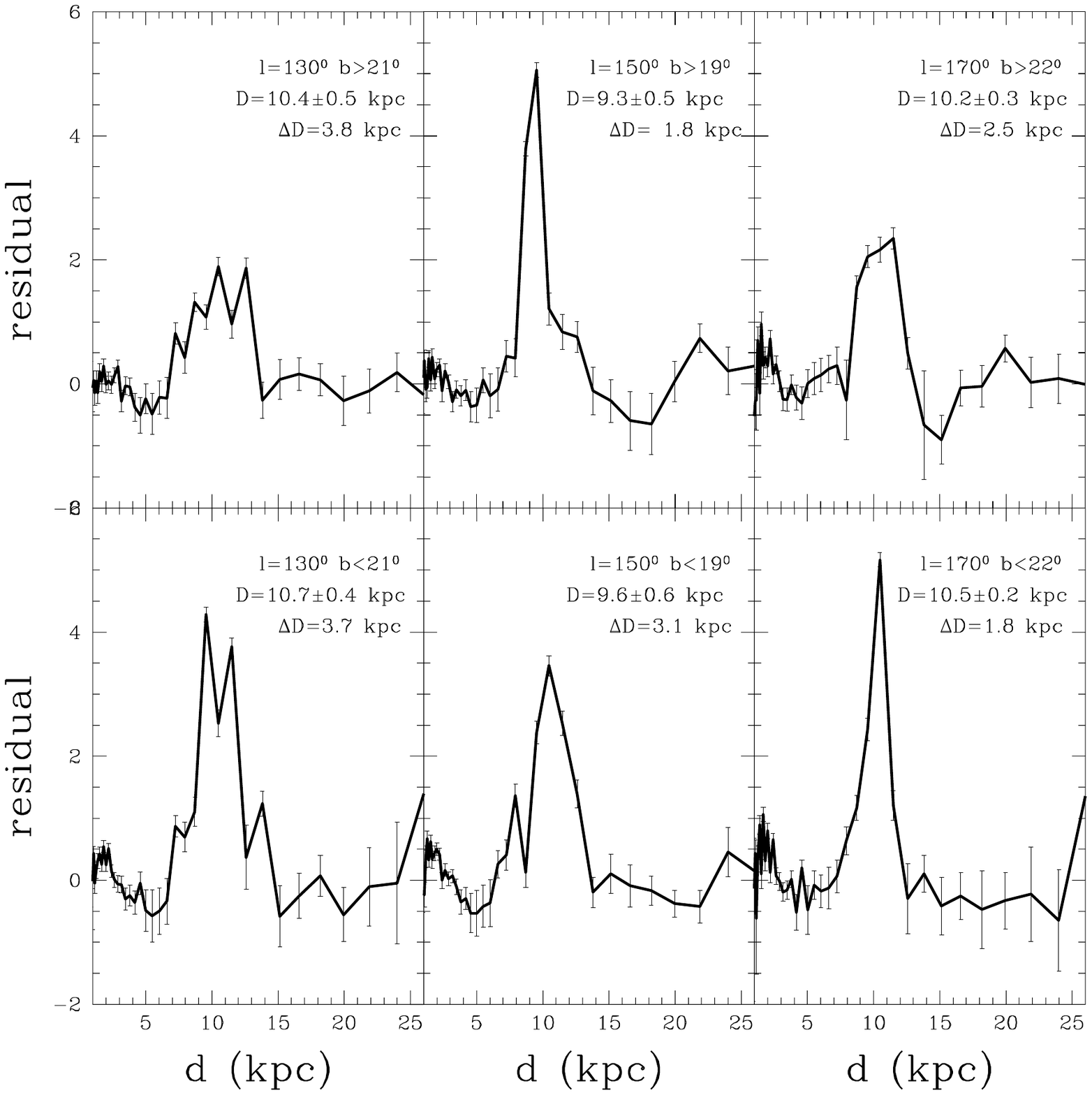}
\caption{The residual density of the stream after subtracting a S\'{e}rsic profile fit, in the distance range of $1 - 25$ kpc, to the smooth Milky Way component. The top panels are the residuals from the higher latitude half of each stripe, while the bottom panels are the residuals from the lower latitude half of the stripes. See Table~\ref{ResultsTable} for the obtained values from this analysis. The distance to each MO detection is determined through a Gaussian fit to the residuals. The width of the MO is the FWHM of that Gaussian fit. This is shown as `D' and `$\Delta$D' in each sub panel. The error on the distance is related to the error on fitting the Gaussian to the residual density profile.\label{figdensity}}
\end{center}
\end{figure}

The depth of the data allows us to unequivocally delineate the density of stars in the MO. In order to obtain a clear detection of the MO and to find out if there are differences with height above the plane, we gathered all the fields corresponding to each stripe into two halves: upper and lower latitude, i.e., six halves in total, two per stripe. This improves the signal to noise of the MO as individual frames do not contain enough stars to perform the analysis. To quantify the resultant overdensities, we removed the smooth background stellar density distribution and fitted a S\'{e}rsic profile to the stellar mass density relation obtained from the CMD analysis. For each density profile, we took all the points within the 3$\sigma$ values of the S\'{e}rsic fit and recalculated the density, thus removing the bulk Milky Way components from the distribution.  The best-fit S\'{e}rsic parameters can be found in Table~\ref{ResultsTable}.The resultant residual for each of the six fields is shown in Figure~\ref{figdensity}.  The upper panels represent the residuals of the total mass density in the upper latitude set and the lower panels are the same for the lower latitude set. All of the residuals show a bump at the location of the MO. The number density is relatively constant across the stripes for each latitude range, however the lower latitudes are consistently denser than the higher latitudes. Figure~\ref{figdensity} also shows the location of the MO, denoted by D, which was found by fitting a Gaussian profile to the residual density peaks. The line-of-sight depth of the MO represents the full width at half maximum of the best fit Gaussian. The stellar number density of each of the six regions can also be found in Table~\ref{ResultsTable}, as well as the mean latitude, heliocentric distance and line-of-sight depth.

\begin{figure}
\epsscale{1.2}
\includegraphics[width=85mm, angle=0]{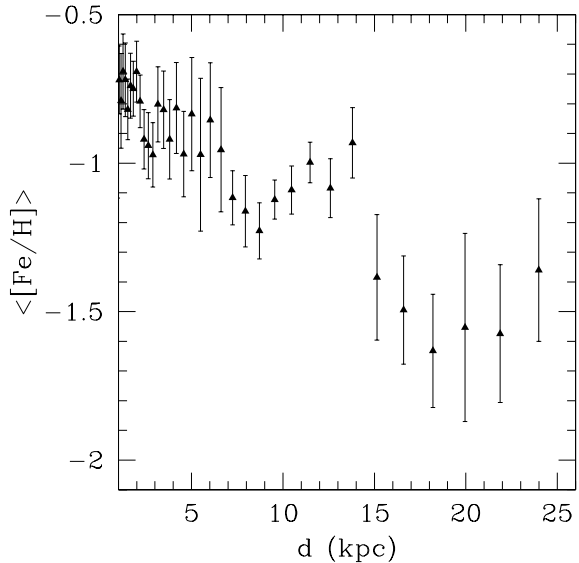}
\caption{Mass-weighted mean metallicity from the MATCH fits, averaged over all latitudes and longitudes.  At distances less than 10 kpc, the metallicity gradient is steep due to the thick disc halo transition. The MO, around 10 kpc, also appears to be a distinct feature in the mean metallicity compared to the `underlying' distribution seen again beyond $\sim$15 kpc. Below 7 kpc, the stellar populations are most likely too complex for our simple approach as seen in the large error bars irregular profile. For a complete discussion on the metallicity of the MO, see Section~\ref{metal}.}\label{figmetals}
\end{figure}

Given the similarities between each of the stripes, we gathered all the stripes together to obtain metallicity profile. The total mass-weighted mean metallicity profile is shown in Figure~\ref{figmetals}. 
Although the smooth underlying Milky Way population has not been subtracted, the MO is still clearly visible. The metallicity distribution is at slightly higher distances than seen in the density profiles and deviates from the smooth background between 9 and 14 kpc heliocentric. It reaches a peak metallicity of [Fe/H]$\sim -$1.0 which is consistent with the photometallicities of the MO as determined through SDSS photometry by \citet{2008ApJ...684..287I}. 

\section{Discussion}\label{discus}
 \begin{figure}
\includegraphics[width=90mm,angle=270]{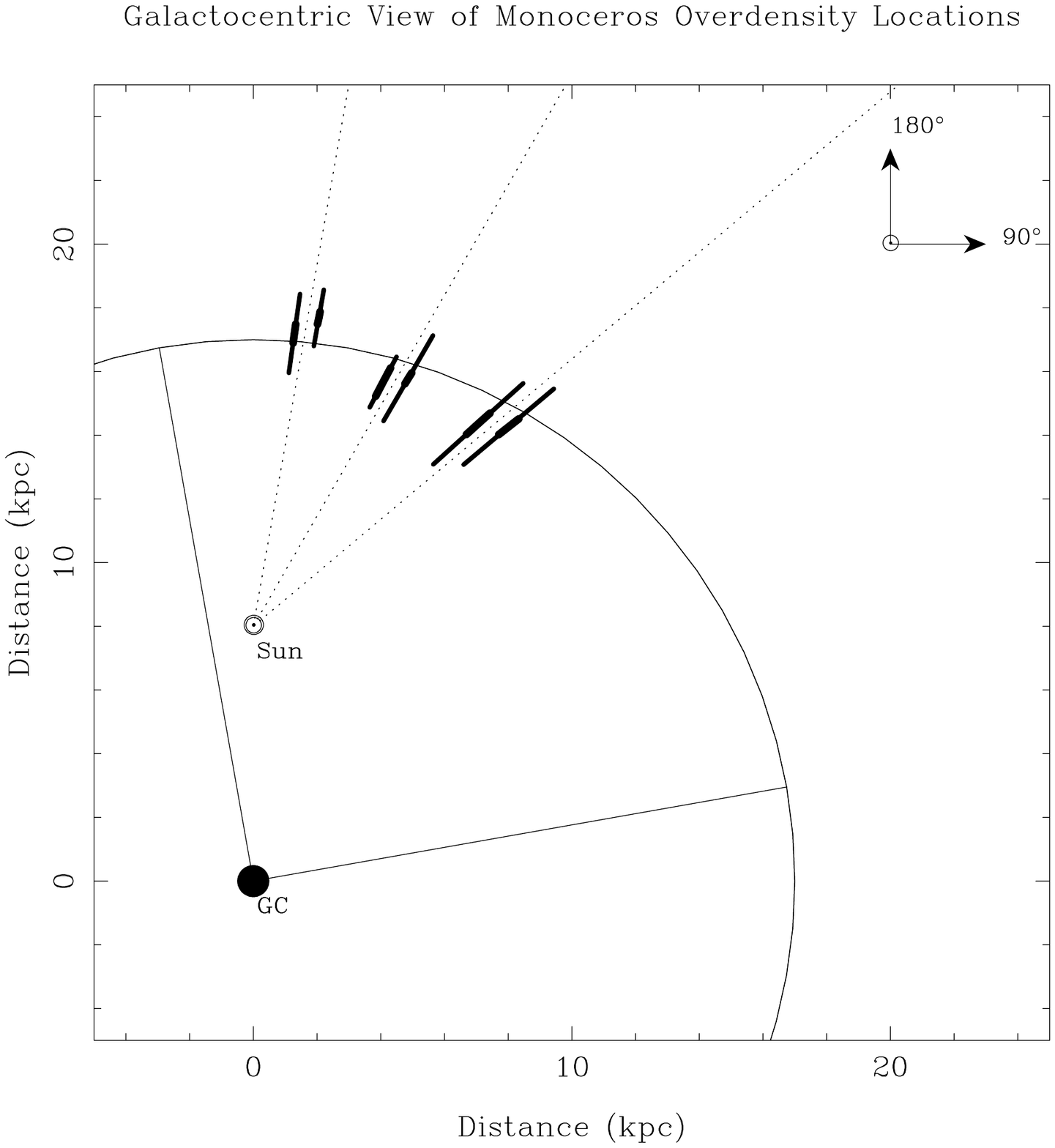}
\caption{Galactocentric locations and distances of the MO in each of the three stripes. The thick line shows the error on the distance estimate while the thin line shows the depth of the stream. The dashed line is the direction of each of the stripes. The lower latitude fields plotted to the right of the line and the higher latitude fields to the left. The Sun is located 8 kpc from the centre and the solid line is a galactocentric circle with radius 17.0 kpc. The lines from the galactic centre are visual aids to highlight the circularity of the curved line.}  \label{figGCzoom}
\end{figure}

These deep CMDs of the MO at  three different galactic longitudes ({\it l} =130$^\circ$, 150$^\circ$, and 170$^\circ$) and covering a range of galactic latitudes (+15$^{\circ}\leq ${\it b}$ \leq$+25$^\circ$) allow us to accurately constrain the structural properties of the MO in these directions. Figures~\ref{figfinalphot},~\ref{figsersic} and ~\ref{figdensity} show that the MO is easily identifiable at all stages of the analysis: it appears as a strong main sequence type feature in the CMDs; it shows a clear excess above the S\'{e}rsic fit to the bulk Milky Way components in the stellar density profiles; occupies a distinct distance range within the sensitivity limits of the method; and has a metallicity that strongly differs from the the background Milky Way population. Figure~\ref{figGCzoom} shows the locations of the MO with respect the Galactic centre and the Sun. A line has been drawn at the Galactic radius of 17.0 kpc for reference. Each detection is shown illustrating its distance uncertainty and the width of the feature. In Sections~\ref{tidal}, \ref{flare} and \ref{disc}, we will discuss the various formation scenarios in light of the density profiles uncovered here. In Section~\ref{metal}, we will discuss the implications of the metallicity finding and its relevance to the outer disc.

 \subsection{The Monoceros Overdensity as a Tidal Stream}\label{tidal}
\begin{figure}
\epsscale{.9}
\includegraphics[width=80mm,angle=270]{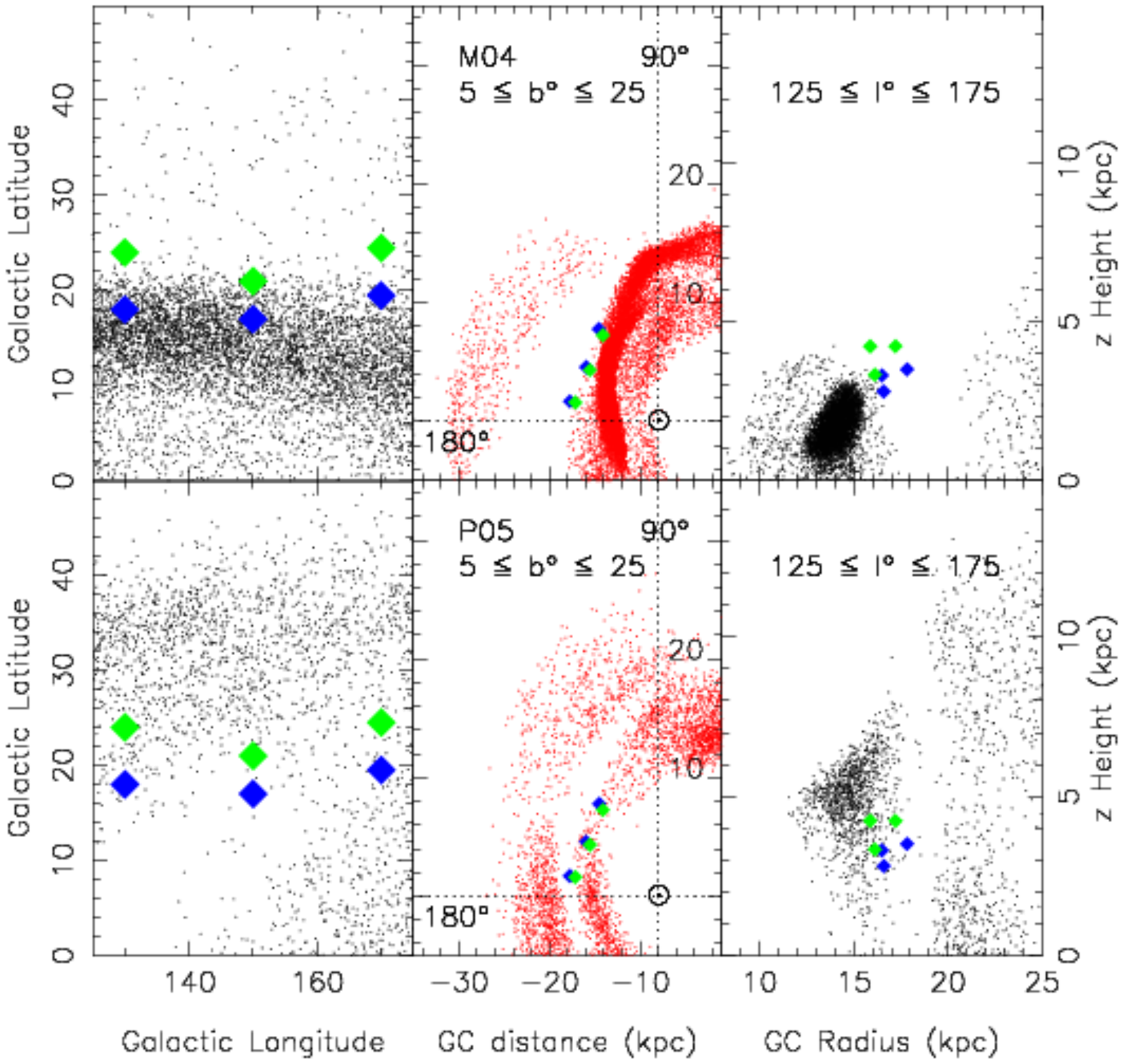}
\caption{Left Panels: \citet{2004MNRAS.348...12M} (top) and ~\citet{2005ApJ...626..128P} (bottom) in the same region as surveyed in this paper. The green diamonds show the mean latitude of the upper half of this survey and the blue diamonds show the mean latitude of the lower half of this survey. Centre Panels: show the top down galactocentric view of the models in the Second Quadrant. The distances to MO are shown with both green and blue filled diamonds. Right Panels: show the cross-section view of the tidal models with galactocentric radius on the x-axis and height above the plane on the y-axis.} \label{figmodeldatacomp}
\end{figure}

We compare our results with the only two current numerical simulations of \citet{2004MNRAS.348...12M} and \citet{2005ApJ...626..128P}.
Figure~\ref{figmodeldatacomp} shows the comparison between the two models in the Galactic latitude range from 5$^\circ$ to 25$^\circ$, as probed by the survey. 
The \citet{2004MNRAS.348...12M} model (upper panels of Figure~\ref{figmodeldatacomp}) shows a slight decrease in Galactic latitude across the survey and describes a distinct stellar stream predominantly below {\it b} = 20$^\circ$. \citet{2005ApJ...626..128P} (lower panels of Figure~\ref{figmodeldatacomp}) has a tidal stream model which is found mostly at higher latitudes. In this manner, we should expect to see a decrease in density at higher latitudes for the \citet{2004MNRAS.348...12M} model and an increase in density for the \citet{2005ApJ...626..128P} model. In the middle upper and lower panels of Figure~\ref{figmodeldatacomp}, is seen that the observations roughly match both models although the measured change in density (see Table~\ref{ResultsTable}) is contrary to both models. In terms of the height above the plane (right panels of Figure~\ref{figmodeldatacomp}) the overdensity seems to be slightly closer at higher latitudes than at lower latitudes, although with the errors it is consistent with a vertical feature. Both models seem to bracket a possible tidal stream scenario for the MO as determined through this survey. Although the structure of the stream seen in the data is not compatible with either simulation it is difficult to exclude a tidal stream solution since the large number of parameters practically ensure a suitable model is likely to be found. 

\subsection{The Monoceros Overdensity as the Galactic Flare}\label{flare}

\begin{figure}
\epsscale{.9}
\includegraphics[width=71mm,angle=270]{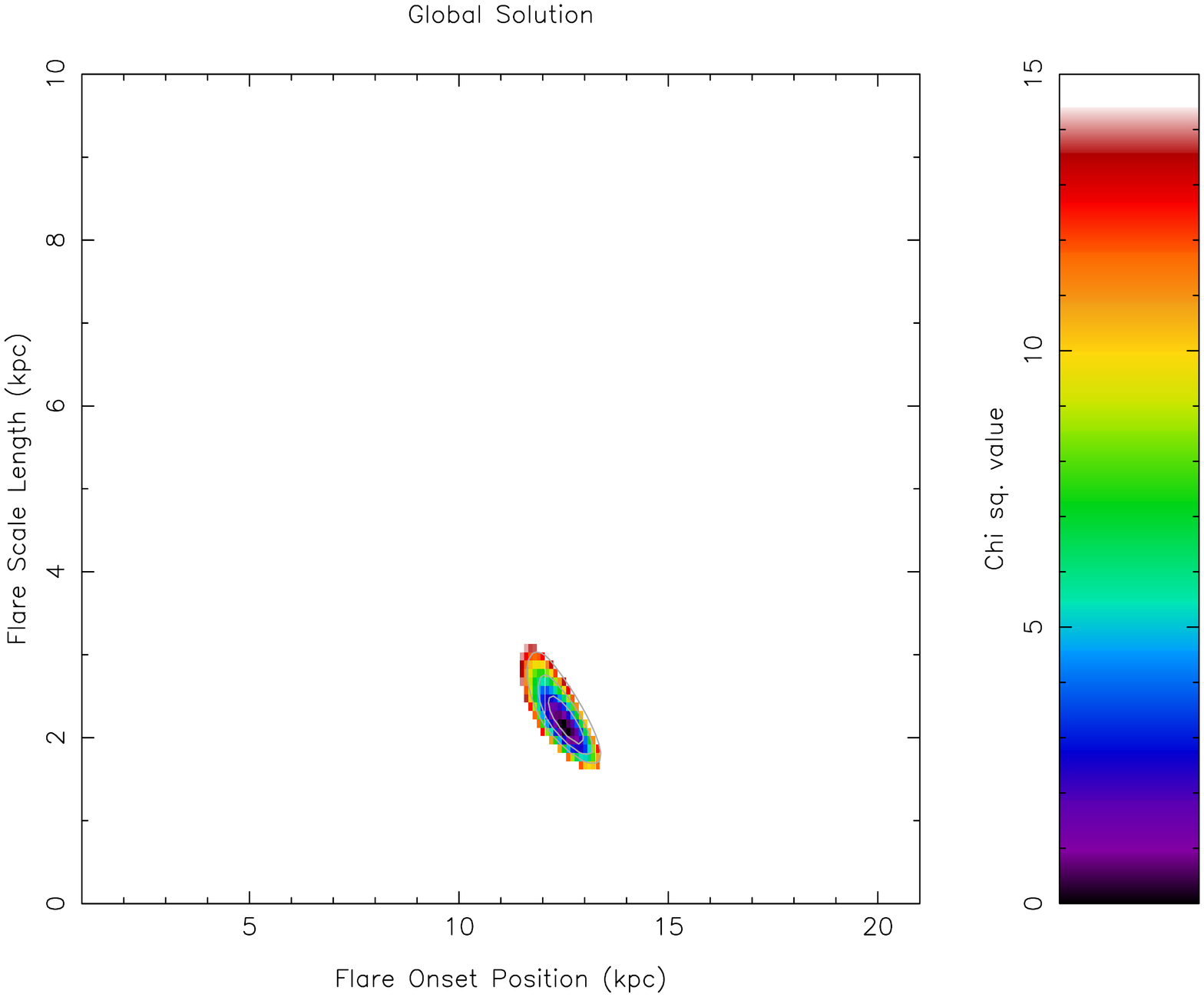}
\caption{Chi-square space after fitting for a global solution to the galactic flare model. The onset position and scale length are free parameters in the fit with the contours showing the 1, 2 and 3$\sigma$ deviations.} \label{figflarechisqglobal}
\end{figure}

\begin{figure*}
\epsscale{.9}
\includegraphics[width=135mm,angle=270]{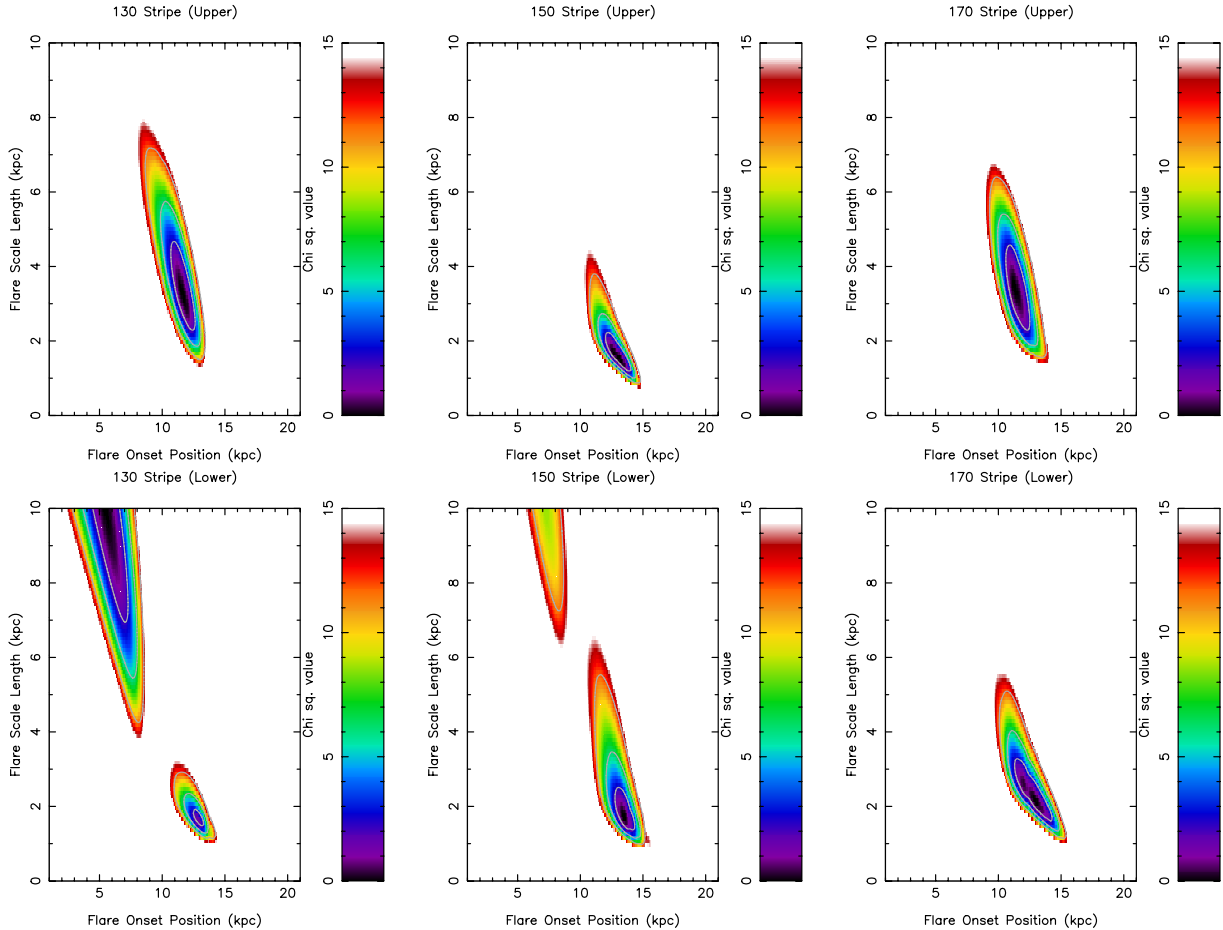}
\caption{Each panel shows the chi-square space after fitting a galactic flare model (Eqn.~\ref{flareeqn}) for the upper and lower halves of the three stripes of the survey with the global solution shown in the top-left panel. The figures show the range of onset points for the galactic flare and its the scale length for the model tested with the resultant chi-square plotted in color. The contours shown in each sub-panel delineate the 1, 2 and 3$\sigma$ lines. In general, the best-fit models are exclusively with very small scale lengths and the onset radii increases from $\sim$12 kpc at {\it l} = 130$^\circ$ to $\sim$14 kpc at {\it l} = 170$^\circ$ inline with the 2 kpc change in distance seen in CMD fitting analysis. The small scale length, in particular, reveals the nature of the MO to be one of a short transient feature and not an extended, generic component of the disc. The minima at onset values less than 10 kpc and large scale lengths seen in the lower halves of the {\it l} = 130$^\circ$ and 150$^\circ$ stripes are due to the model not having a prescription for the warp and the densities mismatching at small heliocentric distances.} \label{figflarechisq}
\end{figure*}

\begin{figure*}
\epsscale{.9}
\includegraphics[width=135mm,angle=270]{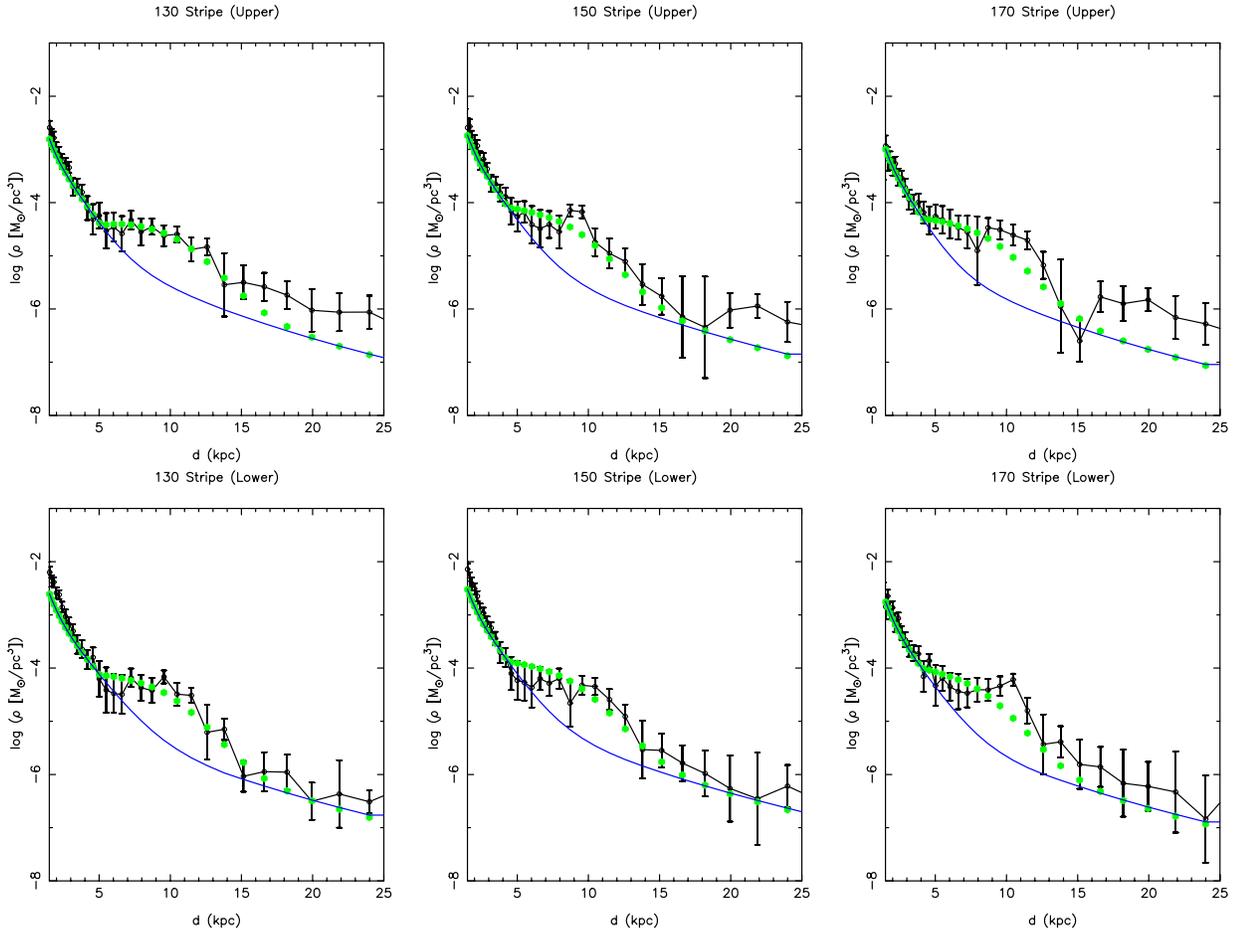}
\caption{CMD fitting analysis density profiles for each of the upper and lower halves of the three stripes of the survey overplotted with the global galactic flare model (green points) plotted against heliocentric distance. The non-flare model is shown as a solid-blue line. A discrepancy within the first 5 kpc is due to the presence of the warp which is unaccounted for by this model and the {\it l} = 130$^\circ$ lower half-stripe has a minima due to this mismatch. Overall, the global flare model is within the error bars although in several fields it does not trace the peak and is consistently lower than the data at large galactic radii.}\label{figflarecomp}
\end{figure*}

The MO is a low-latitude stellar structure and, as such, could be related to the generic structure of the disc. Although many investigations have pursued this possibility  \citep{2004A&A...421L..29M,2006MNRAS.368L..77M,2006A&A...451..515M,2007A&A...472L..47L, 2011A&A...527A...6H}  the distance to the MO typically precludes a definitive conclusion since the MO stars are faint and removing contaminants is highly problematic. Recently, \citet{2011A&A...527A...6H}, have attempted to show that the stellar profiles seen in the directions of the MO are compatible with the flaring of the galactic disc. The flare is described such that beyond a certain radius, the disc rapidly thickens and becomes prominent above the plane, replicating the effect of the MO stars. They sample a small range of galactic longitudes, mostly in regions unaffected by the galactic warp, and fit a small range of flare models to the SDSS CMDs. They conclude that the stellar counts can be accounted for with the galactic flare starting at 16 kpc galactocentric and using a scale length of $\sim$4.5$\pm$1.5 kpc.

Although the CMD fitting method presented here differs significantly from the star count approach used in \citet{2011A&A...527A...6H}, we have fitted their flare models to our dataset across a large parameter space of flare scale lengths and flare onset positions to further investigate this scenario. In this regard, we also use the equations below to define the flare as described in their paper. Table~\ref{flareparams} lists the constants used in the model. For convenience, we reproduce them here (Equation~\ref{flareeqn}), note that $R_i$ is the radius at which the flare starts; $A$ is a scale factor for the density; $\rho_{\text{thin}}$ is the density of the thin disc; $\rho_{\text{thick}}$ is the density of the thick disc; $\rho_{\text{halo}}$ is the density of the halo; $h_{rf}$ is the flare scale length; $R$ is the galactocentric radius and $z$ is the height above the disc.

\begin{deluxetable}{lr}
\tabletypesize{\scriptsize}
\tablecaption{Flare parameters used in Equations ~\ref{eqncomplete} from \citet{2011A&A...527A...6H}.\label{flareparams}}
\tablewidth{0pt}
\tablehead{
\colhead{Parameter} & \colhead{Value}
}
\startdata
thin disc scale height ($h_{z,thin,\odot}$)&186 pc\\
thin disc scale length ($h_{R,thin}$)& 2400 pc\\
thick disc scale height ($h_{z,thick,\odot}$)&631 pc\\
thick disc scale length ($h_{R,thick}$)& 3500 pc\\
Solar radius ($R_\odot$) & 7900 pc\\
\enddata
\end{deluxetable}

\begin{eqnarray} \label{flareeqn}
&\rho_{\text{total}}= \rho_{\text{thin}} + \rho_{\text{thick}} + \rho_{\text{halo}}\\
&\rho_{\text{thin}}\!=\!A\!\left[\,\frac{h_{\text{{\it z},thin,}\odot}}{h_{\text{{\it z},thin}}(\text{\it R})} \right ]\!\text{exp}\!\left [\!-\frac{\text{\it R - R}_{\odot}}{h_{\text{{\it R},thin}}} \right]\!\text{exp}\!\left[\!-\frac{|z|}{h_{\text{{\it z},thin}(\text{\it R})}} \right]\nonumber\\
&\rho_{\text{thick}} = 0.09A\biggl[\frac{h_{\text{{\it z},thick},\odot}}{h_{\text{{\it z},thick}}(\text{\it R})} \biggr]\text{exp}\biggl[-\frac{\text{\it R - R}_{\odot}}{h_{\text{{\it R},thick}}} \biggr]\text{exp}\biggl[-\frac{|z|}{h_{\text{{\it z},thick}(\text{\it R})}} \biggr]\nonumber\\
&\rho_{\text{halo}} = 1.4\times10^{-3}A \frac{\text{exp}\bigl[ 10.093 \bigl (1- \bigl(\frac{R_{sp}}{R_{\odot}}\bigr)^{1/4} \bigr ) \bigr]}{(R_{sp}/R_{\odot})^{7/8}} \nonumber\\
&h_{\text{{\it z},thin/thick}}(\text{\it R}) =  \biggl \{ \nonumber
\begin{array}{l}
	h_{\text{{\it z},thin/thick},\odot}, \text{\it R} \leq \text{R}_{i} \\
 	h_{z,\text{thin/thick},\odot}\text{exp}\bigl(\frac{\text{\it R}-\text{R}_{i}}{h_{r\!f}}\bigr), \text{\it R} > \text{R}_{i}\\
 \end{array}\\
&\text{\it R}_{sp} = \sqrt{\text{\it R}^{2} + 2.52{\text{\it z}}^{2}} \nonumber
\end{eqnarray}

\begin{deluxetable}{lcc}
\tabletypesize{\scriptsize}
\tablecaption{Best Flare Model Parameters determined by the minima of chi-square map as seen in Figure~\ref{figflarechisq} with the uncertainties determined by the 1$\sigma$ contour line. \label{flarebestfits}}
\tablewidth{0pt}
\tablehead{
\colhead{Field} & \colhead{Onset Position (kpc)} & \colhead{Scale Height (kpc)}
}
\startdata
Global & 12.6$^{+0.3}_{-0.6}$ & 2.1$^{+0.3}_{-0.2}$\\
130 (Upper) & 11.7$^{+0.8}_{-1.1}$ & 3.2$^{+1.4}_{-1.0}$\\
130 (Lower)\footnote{The lower field in the 130 stripe has no error estimate since the minima used here is not the absolute minima found in Figure~\ref{figflarechisq}.} & 12.8 & 1.7\\
150 (Upper) & 13.2$^{+0.7}_{-1.3}$ & 1.5$^{+0.7}_{-0.4}$\\
150 (Lower) & 13.5$^{+0.7}_{-0.8}$ & 1.8$^{+0.6}_{-0.5}$\\
170 (Upper) & 11.3$^{+1.1}_{-0.8}$ & 3.5$^{+1.0}_{-1.3}$\\
170 (Lower) & 13.0$^{+1.2}_{-1.8}$ & 2.2$^{+1.0}_{-0.7}$\\
\enddata
\end{deluxetable}

To test this model against our MO density profile we have varied both the onset point of the flare and its scale length. Figures~\ref{figflarechisqglobal} and \ref{figflarechisq}, show the resulting chi-square space with 1, 2 and 3$\sigma$ contours for the global and individual fits to the data, respectively. The flare onset position varies from 1 to 21 kpc and the scale length has been varied from 0 to 10 kpc, both of which were iterated in 100 pc steps. Each model was compared against the data and the chi-square value was determined for each point in the parameter space. Flare models with small scale lengths describe sharp features in the density profile while large scale lengths have long slowly varying density profiles. The data has been fitted both to find a global solution for the flare considering all the data and individually to highlight the differences between fields. Figure~\ref{figflarechisq} shows that the MO density profile typically requires the flare model to have a very short scale length, with a global solution of 2.1 kpc and an onset radius of 12.6 kpc (see Table~\ref{flarebestfits} for the best fit in each field). In Figure~\ref{figflarecomp}, the global model has been overplotted on each of the MO density profiles for comparison with the non-flare model shown as a solid-blue line. The model of the flare used here is very basic and so does not include a prescription for known Galactic features such as the Warp. This is clearly evident in the fields closer to {\it l} = 90$^\circ$ where the Warp becomes stronger. The mismatch at small heliocentric distances for the lower half of {\it l} =  130$^{\circ}$ stripe is an example of this.

In general, this basic flare model can be fit to the data within the uncertainties for the majority of the points. While the presence of the warp is clearly responsible for the discrepancies at small heliocentric distances, it is unclear whether the differences between the MO density profile and the generic properties of the flare model should be explained by the noise in the data or an intrinsic irregularity of the outer disc.  The flare parameters found here are consistent with that found by ~\citet{2011RMxAC..40..245M}  (onset $\sim$ 11.5 kpc, scale length $\sim$1.6 kpc), using RR Lyrae stars to trace the outer thick disc. However,  this is a much shorter scale length and onset radius than that found by \citet{2011A&A...527A...6H} (onset $\sim$ 16 kpc, scale length $\sim$4.5$\pm$1.5 kpc). This difference potentially arises from the fact that  \citet{2011A&A...527A...6H} push to stars with increasing photometric errors which will inherently smear out their result.  Figure~\ref{figflarechisq} shows that a scale length of $\sim$4.5$\pm$1.5 kpc is feasible at the $2 - 3\sigma$ level but a flare onset position of 16 kpc is not likely with our data. Improved number statistics might be necessary to reduce the uncertainties in the MO density profile and thus we could determine whether the deviations from the smooth model can be considered significant or simply the nature of the outer disc itself. 

\subsection{The Monoceros Overdensity as a Perturbed disc}\label{disc}
There is the possibility that the MO could be explained through a disrupted disc scenario whereby the disc interacts with a massive dark matter sub-halo. In this scenario, no new stars are added to the disc but rather the existing disc stars are swept or migrated into large spiral or ring-like structures. Models illustrating this scenario can be found in \citet{2008ApJ...688..254K, 2008ApJ...676L..21Y,2011Natur.477..301P,2011MNRAS.tmp.1860G}. All the authors find that the structures are typically $\sim$4 Gyrs old and so are relatively long-lived. 

Distinguishing between these and a tidal stream is difficult and most likely requires detailed velocities or chemical abundance information. Fortunately, some insights into whether this scenario is feasible can be seen in the stellar density profiles as shown in \cite{2011Natur.477..301P}. In their model, the resulting stellar density profile with heliocentric radius has significant substructure. Since the overdensity is not created with new stars but rather a rearrangement of the disc, it follows that creating an overdensity naturally produces a corresponding underdensity. In this manner, both their light and heavy Sagittarius-like dwarf galaxy encounters induce significant underdensities in the disc, adjacent to the ring-like overdensity of around 1.2 dex (see Fig 4b and S7 from ~\citet{2011Natur.477..301P}). Crucially, there is no evidence for these underdensities in our dataset that could possibly match the dramatic change in the stellar density profile as suggested by their model. Additionally, to place stars at the location of the MO detections $\sim$5 kpc above the plane, is limited to their simulation with a heavy Sagittarius dwarf galaxy. The light version is unable to have such an impact on the scale heights of the disc stars.
 
\subsection{The Metallicity of the Monoceros Overdensity and the Outer Disc}\label{metal} 
The metallicity of the MO has been measured both photometrically and spectroscopically with a variety of results. Photometrically, the metallicity has been found to be [Fe/H]$\sim -0.95\pm0.15$ from \citet{2008ApJ...684..287I}, [Fe/H]$\sim -1.5\pm0.1$ from \citet{2011ApJ...731....4S} and [Fe/H]$\sim -1.0$, this study. Spectroscopically, it has been reported as [Fe/H]$\sim -1.6\pm0.3$ by \citet{2003ApJ...588..824Y}, [Fe/H]$\sim -0.4\pm0.3$ by \citet{2003ApJ...594L.119C},$ <$[Fe/H]$>\sim -1.37\pm0.04$ by \citet{2005ASPC..336..371W} , $-1.04\leq$[Fe/H]$\leq -0.1$ by \citet{2010ApJ...720L...5C}  and most recently by \citet{2012arXiv1205.0807M} with $[Fe/H] = -1.0$. From this, it is clear that the MO is a complex stellar population that is consistently metal-poor.

Characterizing the disc at the distances of the MO is difficult and so the expected metallicity profile needs to be extrapolated from our understanding at smaller galactocentric radii. To do this, we utilize five studies of the outer disc \citep{2012MNRAS.419.2844C,2011ApJ...738..187L,2012ApJ...746..149C,2008ApJ...684..287I,2005AJ....130..597Y} to understand how the disc evolves at these distances. Using the SEGUE survey, \citet{2011ApJ...738..187L} traced the metallicity of the thin and thick discs beyond 2 kpc from the Sun showing that the mean [Fe/H] of the thin disc is $\sim -0.2$ and for the thick disc $\sim -0.6$. \cite{2012ApJ...746..149C} also finds similar results with SEGUE finding that stars between 1.0 - 1.5 kpc above the plane have a metallicity of $-0.3 \leq$[Fe/H]$\leq$-1.0, centered on [Fe/H]$\sim$-0.6.  In terms of metallicity gradients, \citet{2012MNRAS.419.2844C} has shown with the RAVE\footnote{RAdial Velocity Experiment} dwarf stars that while the thin disc decreases in metallicity by $-0.043$ dex/kpc$^{-1}$, the thick disc is essentially flat. Both \citet{2011ApJ...738..187L} and \citet{2012ApJ...746..149C} also find the thick disc to have no metallicity gradient. \citet{2005AJ....130..597Y} explored the outer disc in the third galactic quadrant using open clusters and also found a flat distribution with [Fe/H]$\sim$-0.6.   \citet{2011ApJ...735L..46B} confirm the metallicity gradient in the thin disc as their target stars in the galactocentric distance range of 9 - 13 kpc have thin disc abundance patterns with a mean metallicity of [Fe/H]$\sim -0.48\pm0.12$, which is significantly more metal-poor than the local thin disc stars. Extrapolating the metallicity gradient of $-0.066$ dex/kpc$^{-1}$ from \citet{2012ApJ...746..149C} to the radii of the MO, we find a predicted thin disc metallicity of [Fe/H]$\sim -0.44$, which is still more metal rich than all estimates of the MO metallicity.

A final possibility remains that the metallicity derived through isochrone fitting is wrong, simply because we have utilized old metal poor isochrones which would be unsuitable for a thin disc population. If the MO were thin disc stars then a 4 Gyr isochrone would better represent such a population. In this case, the isochrone would be bluer by $(g_{o} - r_{o})\sim 0.1$ and so could feasibly be consistent with our data. The difficulty with this approach is that a 4 Gyr old main sequence turn-off star is at least 1 magnitude brighter than the corresponding 10 Gyr star. At the turn-off magnitudes seen in the data, $g_{o}\sim 19$, this translates into an additional 5 kpc in line-of-sight distance, placing the MO at $\sim$21.5 kpc Galactocentric. Naively extrapolating the metallicity of the disc to these distances results in [Fe/H]$\sim -0.7$. The MO though, is now 10 kpc above the Plane and so at each turn it becomes harder to associate thin disc stars with the MO. If the thick disc truly exhibits no change in its metallicity distribution with radius then the MO also remains more metal poor than the thick disc at these large radii. 

Given this understanding of the outer disc, we can interpret the likelihood of the different formation scenarios with the metallicity finding of this study. It is important to note that this method of determining the metallicity relies on the bulk properties of the stars in the CMD and is not a direct measure of distinct components like the Thick disc and Halo. Rather, the profile as seen in Figure~\ref{figmetals} shows how the contributions of the Halo stars become more dominant with increasing distance and so the average metallicity of the stellar populations present is increasingly more metal poor with heliocentric distance. The MO is therefore a distinct population which abruptly appears against this smooth transition to a pure Halo population beyond the disc.

{\it Tidal stream scenario}: Distinguishing between local disc stars and stream stars from a merger is perhaps clearest in the chemical abundance patterns as shown in the review by ~\citet{2009ARA&A..47..371T}. There are distinct chemical differences between local MW stars and stars from nearby galaxies that reveal their different enrichment histories. Recent studies of the MO using spectroscopically determined abundances   (e.g. \citet{2010ApJ...720L...5C,2012arXiv1205.0807M}), show that the chemical properties of the MO are closer to a Large Magellanic Cloud or Sagittarius Dwarf galaxy type abundance pattern than a pure Milky Way disc population. This offset in metallicity between the MO and outer disc suggests potentially a different origin for these stars. \citet{2006ApJ...650L..33P} suggest the outer disc could have been created through a series of mergers in which case the abundance pattern and the consistently metal-poor nature of the  MO member stars are supportive of this scenario. Additionally, \citet{2010ApJ...712..692C} discuss the similarities between the metal-weak thick disc (MWTD) and the MO suggesting the two may be related. Indeed the MWTD itself is presented in \citet{2010ApJ...712..692C} as distinct to the canonical thick disc and as such is possibly the result of a merger with the Milky Way disc. Together the evidence builds that the MO is an accretion event although there is no viable progenitor and its passage through the outer disc is still unknown.

{\it Galactic Flare scenario}: Although the flare model of \citet{2011A&A...527A...6H} does not make any predictions about the metallicity of the stars, our understanding of the disc can be used to determine whether the MO metallicity is consistent with a Milky Way population. It is clear from Figure~\ref{figmetals} that the stars along our lines of sight have a steadily declining metallicity with distance and the stars bracketing the MO typically have abundances of $-1.2 <$ [Fe/H]$ < -1.5$. Thus the MO appears distinct in the outer disc as more metal-rich than the nearby stars. A comparison between these metallicities and those described in \citet{2008ApJ...684..287I} suggest that at these distances we are beginning to probe the inner halo prior to MO and beyond the MO there is a clean halo sample. Clearly, these stars are apart from the main disc population but it is difficult to explain why the MO is so metal poor if it is simply an extension of the underlying disc. The flare is undoubtedly a real phenomenon but to what extent and what influence it has in the outer disc is uncertain.

{\it Perturbed disc scenario}: The stars which are perturbed into the MO-like structure seen in  \cite{2011Natur.477..301P} are sourced from across the entire disc. The member stars are migrated from inside and outside of the final location and so the resultant metallicity should be an average of these contributing locations in the disc. Since the disc, in general, is more metal rich than the MO and there are very few locations within the disc which could supply stars more metal poor than the MO, it is highly unlikely that an aggregate population as proposed by this model, could achieve the metal poor status of current set of MO metallicity estimates. Since our findings too, confirm the metal poor nature of the MO, the perturbed disc scenario with its predicted observable properties, as described by \cite{2011Natur.477..301P} is not feasible given the data.

\section{Conclusion}\label{conclusion}

We have presented new distance, density and metallicity measurements for the stellar Monoceros Overdensity (MO) in the outer Milky Way, based on SUPRIME-CAM wide field imaging data and a CMD fitting analysis. Our distance measures are the most quantitative estimates to date for the MO.

The MO appears as a wall of stellar material at roughly 10 kpc from the Sun at the galactic longitudes of 130$^\circ$, 150$^\circ$ and 170$^\circ$, and galactic latitudes of $+15^\circ \leq$ b$\leq +25^\circ$. Detections of the MO have been confirmed between 3 - 5 kpc above the plane and consist of a metal-poor population with an average metallicity of [Fe/H]$\sim -1.0$.

We consider these findings in the light of the three formation scenarios currently in the literature: (i) a tidal stream origin; (ii) the galactic flare; and (iii) the perturbed disc. We find that: 

\begin{enumerate}
\item[(i)] Tidal stream models from the literature bracket the distances and densities we derive for the MO. Furthermore, recent results for the chemistry of stars in the MO support an extragalactic origin. This suggests that a tidal stream model can be found that would fully fit the MO data. On the other hand, the large parameter space available for this model: the orbit, mass, inclination and eccentricity of the merger, amongst others, presents the danger that such a fit - while possible - might not be the true explanation for the MO.

\item[(ii)] The flaring of the galactic disc provides another possibility for explaining the presence of these stars at large distances from the plane. We fitted a large range of galactic flare models finding a solution with a mean onset radius of 12.6 kpc and a scale length of 2.1 kpc that is a reasonable match the data. This is similar to the findings of \citet{2011RMxAC..40..245M} but is much smaller than the models suggested by \citet{2011A&A...527A...6H}. The main difficulties with the flare model are: (a) whether the basic flare model used here while consistent with the data would be applicable across wider latitude and longitude ranges and (b) the metallicity ([Fe/H]) derived in this paper as well as the determinations from other sources (see Section~\ref{metal}) are building a consistent picture that the disc is too metal-rich to source the MO stars. If the disc can be shown to be metal-poor at these radii then the flare scenario is indeed a possibility.

\item[(iii)] The perturbed disc scenario makes clear testable predictions about the metallicity and stellar density profile of a MO-like feature. Both of these are incompatible with the data: the MO stellar density profile does not contain the significant underdensities predicted by the model while the metallicity of the MO is too metal-poor even for a population of stars sourced from across the disc. 

\end{enumerate}

It is clear that the MO still lacks the observational evidence required to unequivocally determine its origins. However, the deep observations we have presented here, coupled with CMD-fitting techniques are able to constrain its properties to much greater precision than has previously been possible. We have ruled out the `perturbed disc scenario' for the MO, and found key problems that must be solved if the MO is to be explained by a flared disc. Given the distance to the MO and the uncertainty over its origin, it is crucial to minimize the photometric errors so as to limit their impact when deriving its properties. Further studies of the MO should include high precision photometry to better constrain the physical dimensions of the MO coupled with high resolution spectroscopy for a detailed abundance analysis. This combined approach seems best suited to unravelling the origin of the MO feature.

\acknowledgments

The authors would like to thank Justin Read for many useful discussions on the implications of this work and the referee for helping to refine the conclusions of this work. BCC would like to thank the Max-Planck Institute for Astronomy and the Alexander von Humboldt Foundation Fellowship under which this project was completed. RRL acknowledges support from the Chilean Center for Astrophysics, FONDAP Nr. 15010003, and from the BASAL Centro de Astrofisica y Tecnologias Afines (CATA) PFB-06/2007. NFM acknowledges funding by Sonderforschungsbereich SFB 881 "The Milky Way System" (subproject A3) of the German Research Foundation (DFG). GFL thanks the Australian Research Council for support through his Future Fellowship (FT100100268) and Discovery Project (DP110100678).The software package {\sc TOPCAT} (http://www.starlink.ac.uk/topcat/) was used extensively in the preparation of this paper.



{\it Facilities:} \facility{SUBARU (SUPRIME-CAM)}

\appendix
\section{Online Data}
The data presented in this paper is being made available online and all issues related to the data can be addressed to conn@mpia-hd.mpg.de. The data consists of all objects extracted from each reduced frame in the survey, galaxies and stars, although poor data has been excised from the final catalogue. Individual reduced frames will be available upon request. The catalogue mostly follows the layout of the Cambridge Astronomical Survey Unit Wide Field Camera Pipeline \citep{2001NewAR..45..105I} with the addition of the Galactic coordinates for each object, Right Ascension and Declination are in J2000.0. 

The descriptors for each column in the catalogue is shown in Table~\ref{datalabels} and an example of the data is shown in Table~\ref{data}. The classification scheme is as follows: Stellar are $-1$, Possible Stellar are $-2$, non-Stellar/Galaxy are $1$, Noise is $0$, Possible non-Stellar/Galaxy is $-3$, Crossmatch problem is $-8$, Saturated object is $-9$.  The entire catalogue contains 3.4 millions objects and the breakdown per filter in the entire catalogue is $\sim$643,000 Stellar, $\sim$550,000 Possible Stellar, $\sim$747,000 non-Stellar/Galaxy, $\sim$411,000 Possible non-Stellar/Galaxy in the $g$-band. The $r$-band has $\sim$625,000 Stellar, $\sim$432,000 Possible Stellar, $\sim$764,000 non-Stellar/Galaxy, $\sim$320,000 Possible non-Stellar/Galaxy.

The data here has been extinction corrected following the standard prescription when the extinction is less than $E(B-V) \leq 0.1$, otherwise the relation from \citet{2000AJ....120.2065B}, as shown in Equaton~\ref{dust}, has been used. The dust values have been extracted from the \citet{1998ApJ...500..525S} maps using the \textsf{dust\_getval.c} program.

\begin{equation}
   	\mbox{dust value} = 0.1 + 0.65(E(B-V) -0.1)
  \label{dust}
\end{equation}

The data has also been corrected for airmass. The airmasses and field centers for each field can be found in Table~\ref{fieldcentres}. Individual objects do not have an identifier which relate them to a particular field however most objects will be easily matched with its corresponding field center. Objects in overlap regions can be associated with a particular field based on the chip in which they reside. In this regard, users should note that duplicate observations of the same object have not been removed from the catalogue. They have been left in to allow the user to gauge the relative depths of each pointing.

\begin{deluxetable}{cllcll}
\tabletypesize{\scriptsize}
\tablecaption{Column information and format. \label{datalabels}}
\tablewidth{0pt}
\tablehead{
\colhead{Column Number}&\colhead{Label}&\colhead{Type}& \colhead{Column Number}&\colhead{Label}&\colhead{Type}
}

\startdata
1 & Right Ascension  (Hours)& Integer& 11 & Y pixel g& Real\\
2 & Right Ascension (Minutes)& Integer& 12 & g mag& Real\\
3 & Right Ascension (Seconds)& Real& 13 & g mag error& Real\\
4 & Declination (Degrees)& Integer& 14 & g classification& Integer\\
5 & Declination (Minutes)& Integer& 15 & X pixel r& Real\\
6 & Declination (Seconds)& Real& 16 & Y pixel r& Real\\
7 & Galactic Longitude ({\it l})& Real& 17 & r mag& Real\\
8 & Galactic Latitude ({\it b})& Real & 18 & r mag error& Real\\
9 & Chip& Integer& 19 & r classification& Integer\\
10 & X pixel (g)& Real& 20 & E(B-V)& Real\\
\enddata
\end{deluxetable}

\begin{deluxetable*}{lllcrrllcr}
\tablecolumns{10}
\tabletypesize{\scriptsize}
\tablecaption{Example data from the catalogue \label{data}}
\tablewidth{0pt}
\tablehead{
\colhead{Col 1.}&\colhead{Col 2.}&\colhead{Col 3.}&\colhead{Col 4.}&\colhead{Col 5.}&\colhead{Col 6.}&\colhead{Col 7.}&\colhead{Col 8.}&\colhead{Col 9.}&\colhead{Col 10.}
}
\startdata
5&32&29.55&62&16&34.7&149.930191&15.2970066&1&1553.4\\
5&32&40.61&62&14&34.4&149.971298&15.2990808&1&1162.8\\
5&32&30.45&62&7&57.5&150.060394&15.2272711&1&1517.3\\
5&32&30.43&62&12&27.5&149.992889&15.2644424&1&1522.6\\
5&32&42.7&62&16&15.0&149.948212&15.3164396&1&1087.9\\
5&33&7.83&62&6&31.6&150.119278&15.2787237&1&184.5\\
5&32&35.22&62&5&13.7&150.106125&15.2127686&1&1343.1\\
5&32&58.89&62&14&42.0&149.987595&15.3309498&1&512.7\\
5&33&6.82&62&6&24.8&150.119965&15.2760792&1&220.4\\
5&32&41.21&62&15&37.5&149.956116&15.3087721&1&1141.0\\[1pt]
\hline\hline\\[-3pt]
Col 11. & Col 12. &Col 13.&Col 14.&Col 15.&Col 16.&Col 17.&Col 18.&Col 19.&Col20.\\[2pt]
\hline\\[2pt]
3848.9&22.552&0.018&-1&1550.4&3851.3&21.162&0.011&-1& 0.113 \\
3248.3&22.56&0.019&-1&1159.7&3250.6&21.157&0.011&-1& 0.145 \\
1251.9&22.555&0.019&1&1514.4&1254.1&20.76&0.0080&1& 0.126 \\
2602.9&22.527&0.019&1&1519.6&2605.2&21.508&0.015&1& 0.102 \\
3758.9&23.611&0.019&0&1084.8&3761.4&3.616&0.0&0& 0.198 \\
840.7&22.586&0.019&1&181.2&842.9&20.898&0.0090&1& 0.122 \\
436.8&22.496&0.019&-1&1340.2&438.9&22.293&0.03&-1& 0.114 \\
3300.5&22.61&0.019&-3&509.4&3302.9&21.231&0.011&1& 0.109 \\
806.1&22.596&0.019&-1&217.2&808.2&21.087&0.01&-1& 0.108\\
3567.9&22.613&0.019&-1&1138.0&3570.4&21.07&0.01&-1& 0.109 \\

\enddata
\end{deluxetable*}

\begin{deluxetable*}{lcccccc}
\tabletypesize{\scriptsize}
\tablecaption{Field centers for the individual pointings of the survey. $g-$ and $r$-band observations have the same pointing centers. \label{fieldcentres}}
\tablewidth{0pt}
\tablehead{
\colhead{Field Name} & \colhead{Galactic Longitude ({\it l})} & \colhead{Galactic Latitude ({\it b})} & \colhead{R.A. (degrees)} &\colhead{Dec (degrees)} &\colhead{Airmass (g)}&\colhead{Airmass (r)}
}
\startdata
130\_01&129.505203&15.2144403&41.34738719353522&76.60936674102940&1.83&1.88\\
130\_02&129.498260&15.5623217&41.99728376377872&76.92390791959453&1.85&1.89\\
130\_03&129.520309&15.9293108&42.83053801772657&77.24082916240711&1.87&1.91\\
130\_04&129.490295&16.3293571&43.57344637209229&77.60785831085641&1.88&1.93\\
130\_05&129.477982&16.7765141&44.54271403857756&78.00559018571801&1.90&1.95\\
130\_06&129.488922&17.2290936&45.68634624732895&78.39321646040258&1.93&1.98\\
130\_07&129.481201&17.7492542&47.01153201618068&78.84294581324289&1.95&2.00\\
130\_08&129.486389&18.2244434&48.36961137572538&79.24196293987690&1.98&2.03\\
130\_10&129.891327&18.6489811&51.37118049084526&79.37541996407468&1.99&2.05\\
130\_11&130.495422&18.4339542&53.12960548181101&78.86147318029894&1.96&2.03\\
130\_12&129.932236&19.6258717&54.88780147544033&80.12706264257046&2.04&2.10\\
130\_13&130.272293&19.6864243&56.53253062677564&79.96761144072845&2.03&2.10\\
130\_14&130.025726&20.5192966&58.82546202768954&80.74218057326074&2.08&2.16\\
130\_15&130.352402&20.5899010&60.48630003355603&80.58115326477444&2.08&2.15\\
130\_16&129.884949&21.1346321&60.96542286637656&81.27529378945327&2.12&2.20\\
130\_20&129.500458&23.1141567&70.27297710747496&82.82702811918269&2.20&2.21\\
130\_21&129.786774&23.3386803&72.92685925574828&82.73512051347623&2.19&2.21\\
130\_22&129.711731&23.7662678&75.57591317062941&83.01743603239895&2.21&2.23\\
130\_23&129.709274&24.1529255&78.39220828275825&83.20574701910283&2.23&2.26\\
130\_24&130.074417&24.3651276&81.19850907262055&83.00000442870187&2.21&2.25\\
130\_25&130.503357&24.6074963&84.23756455977386&82.73757597853790&2.20&2.24\\
130\_26&129.883682&25.2132816&87.35185357260404&83.47482960191130&2.26&2.30\\
130\_27&130.162216&25.6129627&91.31864264785848&83.34836976965003&2.25&2.31\\
130\_28&130.274887&25.7499065&92.67235820452719&83.28326763729932&2.25&2.31\\
130\_29&129.845047&26.3427334&97.17409024268696&83.77700149427237&2.30&2.37\\
150\_01&149.367172&14.7309074&81.47717324443296&62.45923524113733&1.54&1.75\\
150\_02&150.021484&15.0184927&82.70019456402970&62.06101665160047&1.53&1.44\\
150\_03&150.568130&15.1846695&83.56255315448394&61.68566109101802&1.53&1.43\\
150\_04&149.267120&15.8481741&83.48413227595164&63.10354272409283&1.55&1.46\\
150\_05&149.467026&16.1752510&84.32222788603623&63.09321359738084&1.55&1.46\\
150\_06&149.739899&16.3192616&84.87544284777340&62.93090358029885&1.54&1.46\\
150\_07&149.869659&16.5827961&85.51538476758475&62.94394993994977&1.54&1.46\\
150\_08&150.418961&16.8561764&86.57224144450691&62.60087744573428&1.53&1.45\\
150\_09&149.855774&17.7393131&87.79741617372791&63.47297203588319&1.55&1.47\\
150\_10&149.978928&17.9487495&88.33340417316249&63.45655395917793&1.55&1.47\\
150\_11&149.595703&18.8632870&89.88188730710624&64.16482959621780&1.58&1.50\\
150\_12&149.404037&18.9116516&89.81690579089114&64.35037968901068&1.57&1.50\\
150\_13&149.469101&19.4710522&91.06547830802131&64.51266042943159&1.58&1.50\\
150\_14&149.498032&19.8628922&91.93500344111298&64.63475959071147&1.58&1.51\\
150\_15&149.465454&20.0847473&92.39163688508779&64.74451585942646&1.58&1.51\\
150\_16&149.472748&20.4829426&93.27157679386292&64.88004096575455&1.59&1.52\\
150\_17&150.229660&20.8936558&94.73593492322217&64.35456695208589&1.58&1.51\\
150\_18&149.338928&21.2984161&94.99382608269545&65.27239075985462&1.60&1.53\\
150\_19&150.334595&21.2837868&95.66243700857011&64.38920170930041&1.58&1.51\\
150\_20&150.471024&21.6856174&96.63736873754850&64.39421276382994&1.58&1.51\\
150\_21&149.105026&22.5320530&97.67760865552893&65.85253465676418&1.62&1.55\\
150\_22&150.136307&22.7098103&98.72154312222851&64.98692367067588&1.60&1.53\\
170\_05&170.197723&16.5303841&99.33109348057584&45.13093652864079&1.11&1.16\\
170\_06&170.188751&17.0044270&99.95114873787325&45.31547803482419&1.11&1.16\\
170\_07&169.485580&17.7562637&100.6056548374478&46.21387973809730&1.12&1.17\\
170\_08&169.477707&18.0655041&101.0199755684189&46.33078105947673&1.12&1.17\\
170\_09&169.961853&18.3975201&101.7015952165405&46.01629271848925&1.12&1.17\\
170\_10&170.181961&18.6210308&102.1070410738897&45.89765781774504&1.11&1.17\\
170\_11&169.297699&19.4511223&102.8321404576463&46.96444993027531&1.12&1.18\\
170\_12&169.478622&19.8252201&103.4316191531833&46.92585236294776&1.12&1.18\\
170\_13&169.431854&20.2172241&103.9554434974554&47.09324064970371&1.13&1.18\\
170\_14&169.017380&20.3716507&103.9902628410851&47.51086788331664&1.13&1.18\\
170\_15&169.785507&20.5863171&104.6219604394106&46.89467570845383&1.12&1.18\\
170\_16&170.062866&21.1207714&105.4821103329919&46.81214457363338&1.12&1.18\\
170\_17&169.436630&21.6631756&105.9884280355083&47.53028736542868&1.13&1.18\\
170\_18&170.377396&21.3885803&105.9831545077192&46.61307929154813&1.12&1.17\\
170\_19&170.731049&22.1850262&107.2307431939022&46.53162776193600&1.12&1.17\\
170\_20&169.043121&22.7703648&107.4179425403126&48.19269113131208&1.14&1.19\\
170\_21&169.394073&22.6403027&107.3635388673966&47.84569174336261&1.14&1.18\\
170\_22&170.449234&22.9586735&108.2095068314083&46.99719888558490&1.13&1.17\\
170\_23&170.113739&24.0381660&109.6265273415698&47.57695394709530&1.13&1.18\\
170\_24&170.010727&24.2990913&109.9668119381534&47.73308269329995&1.14&1.18\\
170\_25&169.622864&24.6615162&110.3634621605747&48.16363440273511&1.14&1.19\\
170\_26&169.925446&25.1457653&111.1647343607107&48.01089193383915&1.14&1.19\\
170\_27&170.744507&25.3222599&111.6741012150978&47.33086267226461&1.13&1.18\\
170\_28&170.499222&25.7041969&112.1500983351095&47.63318062885149&1.14&1.18\\
170\_29&169.509598&26.5984821&113.1763707919120&48.69161945583298&1.15&1.20\\
170\_30&169.204178&27.2056103&114.0004198180252&49.07784152319226&1.15&1.20\\
170\_31&169.042969&27.5770073&114.5180484244156&49.28747251547379&1.15&1.20\\
170\_32&170.136215&27.5013657&114.6775481523616&48.32084412581273&1.14&1.19\\
\enddata
\end{deluxetable*}


\end{document}